\documentclass[aps,prd,twocolumn,groupedaddress,showpacs,showkeys,preprintnumbers,amsmath,amssymb]{revtex4}
\usepackage{graphicx}
\usepackage{dcolumn}
\usepackage{bm}

\usepackage{psfrag}
\usepackage{subfigure}

\begin{document}
\newcommand{\be}{\begin{equation}}
\newcommand{\ee}{\end{equation}}

\preprint{APS/123-QED}

\title{Exploring Parameter Constraints on Quintessential Dark Energy: the Inverse Power Law Model}


\author{Mark Yashar, Brandon Bozek, Augusta Abrahamse, Andreas Albrecht, and Michael Barnard}
\affiliation{Department of Physics, University of California at Davis, One Shields Avenue, Davis, CA 95616}


\date{\today}

\begin{abstract}
We report on the results of a Markov Chain Monte Carlo (MCMC) analysis
of an inverse power law (IPL) quintessence model using the Dark Energy
Task Force (DETF) simulated data sets as a representation of future
dark energy experiments. We generate simulated data sets for a
$\Lambda CDM$ background cosmology as well as a case where the dark
energy is provided by a specific IPL fiducial model and present our
results in the form of likelihood contours generated by these two
background cosmologies. We find that the relative constraining power of the
various DETF data sets on the IPL model parameters is broadly
equivalent to the DETF results for the $w_{0}-w_{a}$  parameterization
of dark energy. Finally, we gauge the power of DETF ``Stage 4'' data
by demonstrating a specific IPL model which, if realized in the
universe, would allow Stage 4 data to exclude a cosmological constant
at better than the 3$\sigma$ level.  
\end{abstract}


\maketitle

\section{\label{sec:Sec1}Introduction}

A host of cosmological measurements indicate that the universe is
undergoing a phase of accelerated 
expansion. This has been generally attributed to a significant
component of smooth energy with a large negative pressure, referred to
as dark energy (DE) and characterized by an equation of state
parameter $w \equiv \frac{p}{\rho}$. Current measurements
indicate that about 70\% of the density of the
universe today is comprised of this dark energy. Candidates for DE
include a cosmological constant $\Lambda$, and a slowly evolving
dynamical scalar field such as quintessence \cite{ratra88}. In
quintessence models, the cosmic acceleration is driven by a scalar
field $\phi$ slowly evolving in some potential $V(\phi)$. In this
scenario, the parameters of the potential $V(\phi)$ determine the
properties of the dark energy.

In general all DE models have serious unresolved theoretical
problems, and one can make the case in different ways as to which types,
if any (i.e. $\Lambda$ or quintessence DE), are best
motivated \cite{albrecht2007,bousso2008}.  This paper is motivated by
the fact that scalar field quintessence is definitely part of the
theoretical discussion, and thus it should also be part of the process
whereby we evaluate future dark energy experiments.  This paper is 5th
in a series of papers motivated in this way \cite{abrahamse2007,bozek2007,barnard2007,barnard2008}. The
IPL model we consider here is one of the more popular quintessence
models.  One of its attractive features is its ``tracking'' behavior
that make its predictions independent of the initial conditions for
$\phi$, assuming that $\phi$ starts out in the (rather broad) basin of
attraction for tracking.  Also, the behavior of the equation of state
in the IPL model tends to be quite different than for the models
considered in our previous work (see \cite{barnard2008} for a
unified discussion), so this makes it an interesting complement to our other
work. 

Recently, the Dark Energy Task Force (DETF) produced a report that considered the
impact of various projected data sets (referred to as ``data models''
and representing future DE observations) on cosmological parameters in
a standard $\Lambda$CDM cosmological model using the 
``$w_0 - w_a$'' parameterization of the dark energy equation of state
\cite{DETF06}, $w(a) = w_0 + w_a(1-a)$, where the scale
factor $a = 1$ today \cite{linder2002}. They assessed the
impact of a given data set using a ``Figure of Merit'' (FOM), defined as the inverse of the
area inside the 95\% confidence contour in the $w_0 - w_a$ plane for a
fiducial $\Lambda$CDM model.  However, as has been pointed out by a number of authors (e.g.,
\cite{liddle2006}), the two- parameter $w_0 - w_a$ phenomenological
model is not motivated by an actual physical
model of dark energy and exhibits very different behavior compared with
popular dark energy models. Our work (represented by this and our
companion papers \cite{abrahamse2007,bozek2007,barnard2007}) supplements the work of the DETF by assessing
the capability of future experiments to constrain DE
by using an equation of state parameterization that \textit{is}
motivated by a physical model of DE - the well-known inverse power law
(IPL) or ``Ratra-Peebles'' (RP) quintessence model.  This potential
has its own motivations, and is also included here because it
generates a family of functions $w(a)$ that are quite different than
those considered in our other work.

This paper is organized as follows. In Section \ref{sec:Sec2} we
describe the features of the IPL quintessence model and its tracking
properties. While most of the focus of this paper is on the tracking
behavior of the IPL model, we also briefly discuss the non-tracking
transient and ``thawing'' behaviors of this model. In Section
\ref{sec:Sec3} we describe how we parametrize the IPL model for our
MCMC analysis. In Section \ref{sec:Sec4a} we present our MCMC analysis
and results using data forecast by DETF to constrain the IPL
quintessence model around a fiducial $\Lambda$CDM model. In Section
\ref{sec:Sec4b} we give our MCMC analysis for
simulated data generated from a fiducial IPL model.  This allows us to further ascertain
how sensitive future observations may be to deviations from a
cosmological constant and to assess to what extent we can exclude the
$\Lambda$ model if IPL quintessence occurs in nature. In Section
\ref{sec:Sec4c} we briefly discuss our MCMC analysis of non-tracking
regions of parameter space. Finally, we discuss our results and
present our  conclusions in Section \ref{sec:Sec5}. 

\section{\label{sec:Sec2}Tracking Quintessence}
    
For a homogeneous scalar field in an FRW universe, the evolution of
the scalar field, given by its equation of motion, is described by
the Klein-Gordon equation \be \ddot{\phi}+ 3H\dot{\phi}+
\frac{dV}{d\phi} = 0 \label{one}\ee where the Hubble parameter $H$ is
given by the Friedmann equation (with $\phi$ and spatial curvature
also taken into account here) \be H^2 =
\left(\frac{\dot{a}}{a}\right)^2 = \frac{1}{3
  M_P^2}\left(\rho_r+\rho_m+\rho_\phi\right)-\frac{k}{a^2}
\label{two},\ee where $a$ is the scale factor, $M_P \equiv
{8\pi G}^{-\frac{1}{2}}$ is the reduced Planck mass, $\rho_{r}(a)$ is the
radiation background energy density, $\rho_{m}(a)$ is the matter
background energy density, $\rho_{\phi}(a)$ is the scalar field energy
density, and $k$ is the curvature constant. The energy
density and pressure of the scalar field are  \be \rho_{\phi} =
\frac{1}{2}\dot{\phi}^2 + V(\phi) \label{three},\ee \be p_{\phi} =
\frac{1}{2}\dot{\phi}^2 - V(\phi)\label{four},\ee 
where the dots denote derivatives with respect to
time. Eq.~(\ref{one})-Eq.~(\ref{four}) enable us to solve for the
background evolution in a quintessence cosmology, once the potential
$V(\phi)$ and energy densities of the different components, $\rho_m$,
$\rho_r$, etc., have been assigned. If the scalar field rolls slowly
enough such that the kinetic energy density is much less than the
potential energy density, i.e., the slow-roll limit, $\dot{\phi}^2 <<
V(\phi)$, then the pressure $p_\phi$ of the scalar field will become
negative and the field energy will approximate the effect of a
cosmological constant. This indicates that a flat potential $V(\phi)$
is required to give rise to accelerated expansion
\cite{copeland2006}. This slow-roll limit corresponds to $w_{\phi} =
-1$ and $\rho_\phi = const$. It also follows that the equation of
state of quintessence is bounded in the range $-1< w_\phi <1$ and is
usually non-constant. In these models, the dark energy behaves as a
perfect fluid in which the equation of state \be w \equiv
\frac{p_{\phi}}{\rho_{\phi}} =
\frac{\frac{1}{2}\dot{\phi}^2-V(\phi)}{\frac{1}{2}\dot{\phi}^2 +
  V(\phi)},\label{eq-wphi} \ee changes with time and is typically
negative when $V(\phi)$ is sufficiently dominant, as expected during the
recent epoch of accelerated expansion. We  can see from
Eq.~(\ref{eq-wphi}) that $\dot{\phi}= 0$ corresponds to the limit in
which the scalar field is a cosmological constant with  $w_{\Lambda} =
-1$. 

\subsection{\label{sec:Sec2a}Tracking Solutions and behaviors}
It has been demonstrated \cite{zlatev1998,steinhardt-1999-59} that a
subclass of quintessence potentials, including the IPL potential, have
several desirable properties. These include the fact
that the equation of motion of these quintessence models have
attractor-like solutions in the space of trajectories of $\phi$
(called ``tracking'' solutions). A broad set of
initial conditions $\phi_{I}$ and $\dot{\phi_{I}}$ in the early
universe (referred to as a ''basin of attraction'')
evolve toward a common
attractor solution giving the same late time evolution of $\phi$, and thus
allowing the scalar field to induce the present phase of accelerated
cosmic expansion starting from a large range of initial
conditions. The tracking solutions are
characterized by an almost constant $w_\phi$, constrained by $ -1 <
w_\phi < w_B$, where $w_B$ is the equation of state of the
dominating background fluid component. The tracking behavior allows
the value of the accelerating matter density today to be determined by
parameters in the quintessence potential, largely independent of the
scalar field initial conditions \cite{liddle1998}. We note, however, that
although this behavior may help to explain why the dark energy has
come to dominate in recent times rather than some earlier epoch, it
does not solve the ``cosmological constant problem'', especially as it
relates to the zero point energy of the quantum vacuum. 

In \cite{steinhardt-1999-59}, a function \be \Gamma \equiv
\frac{V''V}{(V')^2}\label{gamma1} \ee (where the primes denote
derivatives with respect to $\phi$) was defined for determining
whether a particular potential admits tracker solutions. It was shown
that tracking behavior occurs when either of the following two
conditions are met: (a) $\Gamma > \frac{5}{6}$ , $w_{\phi} < w_B$ ,
$\Gamma \approx const$, (and thus $\left|\frac{V'}{V}\right|$
decreases as V decreases); or (b) $\Gamma <  1$, $\frac{1}{2}(1+w_B) >
w_{\phi} > w_B$, $\Gamma \approx const$, (and thus
$\left|\frac{V'}{V}\right|$ is strictly increasing as $V$
decreases). The only constraint on the initial energy density in the
tracker is that it be less than or equal to $\rho_{B,I}$, the initial
energy density of the background fluid component (matter or
radiation), and greater than $\rho_{m,0}$, the current matter energy
density. This condition is necessary in order for $\phi$ to converge
to the tracker solution before the present time
\cite{steinhardt-1999-59,steinhardt05}. On the other hand, solutions
of the Klein Gordon equation do not converge to tracker solutions for
potentials in which $w_{\phi} < w_B$ and $\left|\frac{V'}{V}\right|$
strictly increases as $V$ decreases ($\Gamma < 1$), or, equivalently,
when $\Gamma < 1-\frac{(1-w_B)}{6+2w_B}$. Note that
$\left|\frac{V'}{V}\right|$ gives the slope of the potential. The quantity
$\frac{V'}{V}$ is also known as a ``slow-roll parameter'' (e.g., \cite{linder2002}) which 
relates to how fast the field moves in the potential for so-called
``slow roll'' solutions. One upshot of the above analysis is that one
can see that potentials (such as IPL) tend not to
have tracking solutions when and where they are flat (that is where
$V''=V'=0$). 

\subsection{\label{sec:Sec2b}The Inverse Power Law Potential}

One of the earliest proposed, simplest, and most widely investigated
of the scalar field quintessence models is the pure inverse power law
(IPL) model, originally introduced by Ratra and Peebles
\cite{ratra88}. This model was originally put forward to mimic a
time-varying cosmological constant undergoing dissipationless decay
and is motivated by supersymmetric QCD (see \cite{masiero2000} and
references therein). More recently, this potential has been reanalyzed
(\cite{steinhardt-1999-59, zlatev1998}) in the context of a scalar
field potential driving the current epoch of cosmic acceleration. 

The IPL scalar field potential is self-interacting, minimally coupled to
gravity, and given by 
\be 
V = V_0 (\frac{M_P}{\phi})^{\alpha}
\label{vipl}.
\ee 
Values of $V_0$ of order the critical density $\rho_c =  3 H_{0}^{2} M_{P}^2$ 
and $\alpha = O(1)$ yield cosmological solutions in which the scalar field can
account for the observed cosmic acceleration today (and typically has
current values $\phi = O(M_{P})$). Furthermore, a large range 
of cosmologically realistic solutions exhibit ``tracking'' behavior
whereby, after some initial transient period, many different solutions
lock on to the same attractor solution. This causes the initial
conditions for $\phi$ to be irrelevant for 
predicting observable cosmological features and removes the need for
tuning of initial conditions seen in many other quintessence models.

It has been shown that the following relation is maintained on the
attractor solutions \cite{ratra88,peebles87,zlatev1998}:  
\be  
\frac{d^2V}{d\phi^2} =
\frac{9}{2}\frac{\alpha(1+\alpha)}{\alpha} (1-w_{\phi}^2)
H^2. \label{vtrack} 
\ee 
The second derivative of the potential gives the scalar field mass
which today is given by $m_{\phi} = V''(\phi_0)
\approx \frac{\rho_{\phi}}{\phi^{2}}$. The tiny value of this mass
($m_{\phi} \sim 10^{-33}eV$) is due to the requirements that $V(\phi)$
slowly varies with the field value and that the current value of
$V(\phi)$ be consistent with observations \cite{peebles87}. When the
scalar field potential is about to dominate we have using Friedman's
equation, $H^2 \sim \frac{V}{M_P^2}$. Then, if $w_{\phi}$ and
$\alpha$ are of order unity, Eq.~(\ref{vtrack}) indicates that the
value of the quintessence field at the present time is of order of the
Planck mass\cite {martin2008}.  

The power law index $\alpha > 0$ determines the shape of the potential as well as the value of $w_\phi$ today.
The slope and curvature of the IPL potential are given by
\be \frac{dV}{d\phi} = -\frac{\alpha}{\phi} V(\phi), \label{slope} \ee
and 
\be  \frac{d^2V}{d\phi^2} = \frac{\alpha(1+\alpha)}{\phi^2} V(\phi). \label{curvature} \ee
We can see that smaller $\alpha$'s  lead to a more flat potential
which will in turn lead to more slowly evolving behavior for $\phi$
(and thus values of $w_\phi$ closer to $-1$). 
Larger values  of $\alpha$ lead to a steeper potential slope, causing more evolution for $\phi$ and
its energy density and also values of $w_\phi$ larger than $-1$. 

Smaller values of $\phi_I$ as well as larger values of $\alpha$ lead to a steeper initial potential slope
and larger values of $V(\phi_I)$. This means that
the scalar field will start rolling from higher up on the potential
and will roll faster, even for cases where the dark energy is initially dominant and $\alpha$ is correspondingly
large, leading to greater evolution of the dark energy 
density.   The quantities $V_0$ and $\alpha$ are the
two free parameters in the potential. In some
supersymmetric QCD realizations of the IPL model \cite{masiero2000},
$\alpha$ is also related to the number of flavors and colors, and can
take on a continuous range of values $\alpha > 0$
\cite{caldwell2004}.  For $\alpha
\rightarrow \infty$ (but with $\rho_\phi$ still subdominant), the scalar field energy density scales like
that of the dominant background. Potentials of
this type also possess the following phenomenological property: they
yield $w_{\phi}$ values which automatically decrease to negative
values at the beginning of matter domination
\cite{eriksson2002}. Given that the energy density of each component
evolves as 
\be 
\rho_i \propto a^{-3(1+w_i)}\label{rhoi},
\ee 
(with $i$ standing for the radiation, matter, or scalar field
component), quintessence will eventually come to dominate the universe
even if it begins as a subdominant constituent.  

The IPL potential is one of a large class of quintessence models
with what has been referred to as ``runaway scalar fields''
\cite{zlatev1998,steinhardt05} whose tracker solutions begin from
some initial $\phi_I$ and $\dot{\phi_I}$ and share some of the
following general features: The field rapidly converges to a point
on the potential where $V'' \approx H^2$, where the Hubble
parameter H is determined by $\rho_m$ and $\rho_r$. As the universe
expands and H decreases, $\phi$ moves down the potential so as to
maintain the condition $V'' \approx H^2$. The universe enters a
tracking phase where $\rho_{\phi}$ catches up to the background
density  $\rho_B$ when
$m_{\phi}^2$ decreases to of order $H^2$ and so $\phi_0 \sim M_P$
\cite{steinhardt-1999-59,copeland2006}. Thus, the distinctive
feature of these tracker fields is that the evolution of the scalar
fields is controlled by $\rho_m$ and $\rho_r$ rather than evolving
independently according to its own potential. This controlled
evolution continues until $\phi$ finally surpasses the point where
critical damping via Hubble expansion occurs. Then the field's own
potential energy is sufficient to freeze the field and cause
$\rho_{\phi}$  to eventually overtake $\rho_m$ and $\rho_r$, driving
the universe into a phase of cosmic acceleration. 

\begin{figure}[h]
\centerline{ 
\scalebox{0.22}{\includegraphics{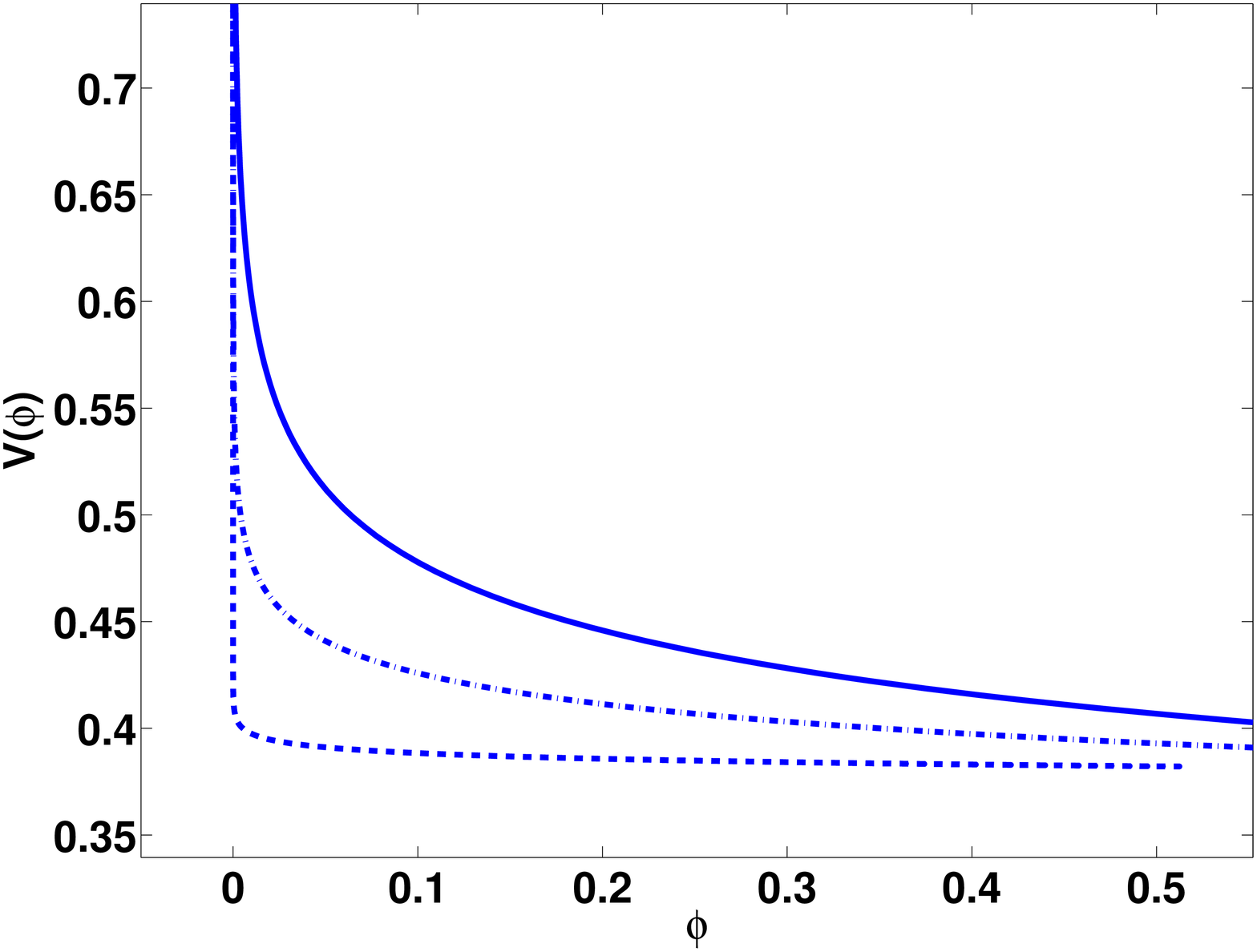}}
}
\vfill
\centerline{
\scalebox{0.22}{\includegraphics{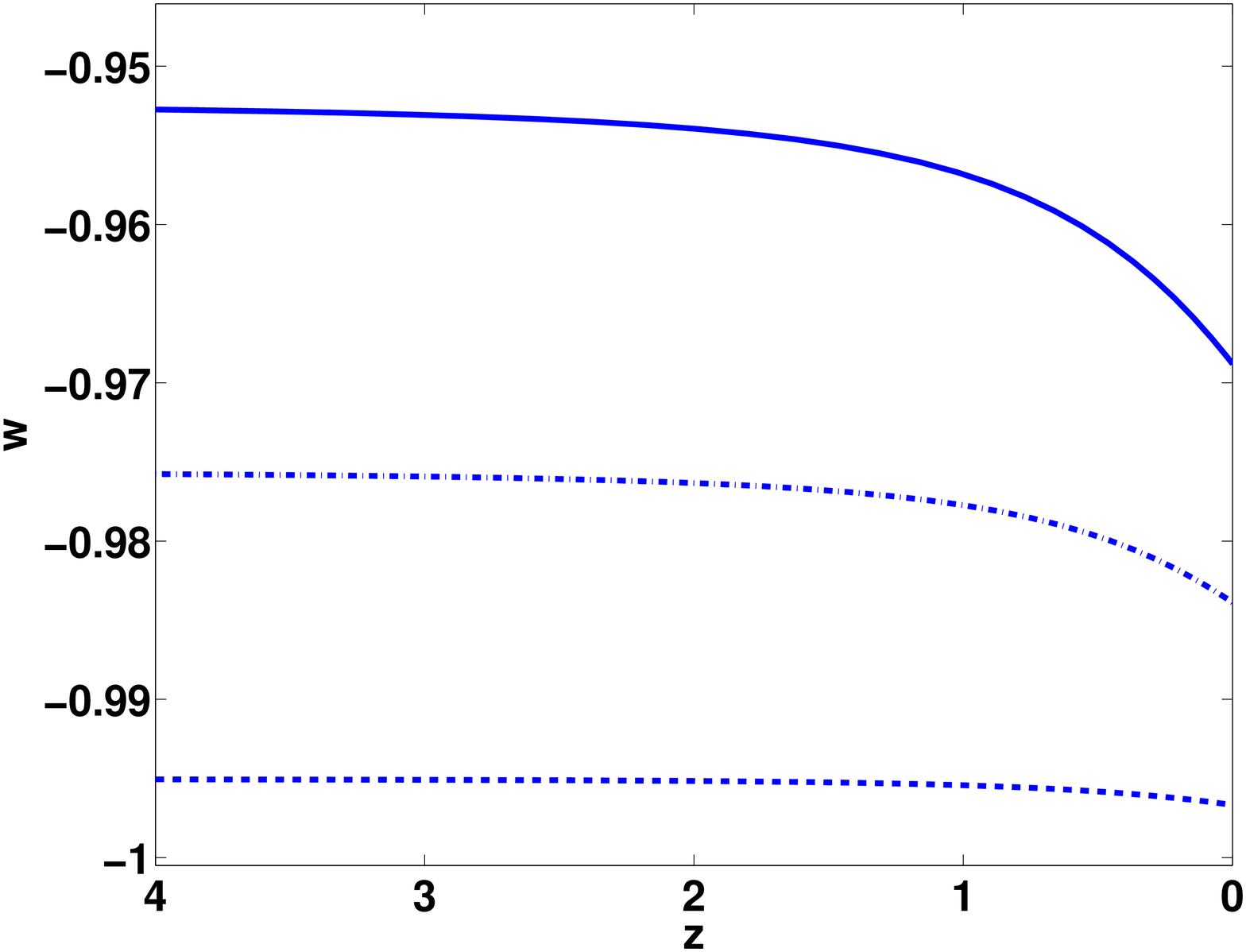}}
}
\caption{\label{fig:fig1ab_ipl}IPL potentials (top panel) and $w(z)$
  evolution (lower panel) for different $\alpha$ values
  (dashed-dotted: $\alpha = 0.05$, dashed: $\alpha = 0.01$ and solid:
  $\alpha = 0.1$). For all curves $V_{0} = 0.38$ and $\phi_I =
  10^{-30}$. Smaller values of $\alpha$ lead
  to flatter potentials and smaller $V(\phi)$. } 
\end{figure}

Figure \ref{fig:fig1ab_ipl} illustrates how the shape of the IPL
potential is changed by selecting four 
different $\alpha$ values for a fixed $V_{0}$.
The value of $\phi_I$ determines where on the potential the scalar
field starts to evolve. 
The present field value $\phi_0$, of order of the Planck mass
$M_P$, is reached from a broad range of initial conditions $\phi_I$
and $\dot{\phi_I}$, with the only important condition being that
$\phi_I << M_P$ \cite{giovi2003}, as consistent with the discussion
concerning tracking in Section \ref{sec:Sec2a} and the more detailed
discussion and criteria  regarding attractor solutions given in
\cite{zlatev1998,steinhardt-1999-59}.  The lower panel of
Fig. \ref{fig:fig1ab_ipl} shows the corresponding evolution of the
equation of state.  For fixed values of $\phi_I$ and $V_{0}$, we see
that larger values of $\alpha$ correspond to $w$ curves with larger
amplitudes and which have larger values today, i.e., deviate more from
a cosmological constant ($w=-1$) at the present time. As $\alpha
\rightarrow 0$, the equation of state more and more mimics the
behavior of $\Lambda$ at late times with $w \rightarrow -1$. The IPL
model has been categorized by \cite{caldwell2005} as a ``cooling'' or
``freezing'' model in which $w>-1$ initially but with $w$ then
decreasing towards $-1$ as the scalar field rolls down the potential. 

For cases in which radiation or matter are dominant and the
contribution of $\rho_{\phi}$ to the expansion of the universe is
neglected, the Klein-Gordon equation gives exact tracking solutions
for the evolution of $\phi$ for the IPL model, as well as the
following time-independent relations between  $\Gamma$, the power law
index $\alpha$ and the equation of state parameter
\cite{zlatev1998,steinhardt-1999-59}, 
\be
 w_\phi = \frac{w_B - 2(\Gamma-1)}{1+2(\Gamma-1)} = \frac{\alpha w_B -
   2}{\alpha + 2}\label{eq-wb},
\ee
where $\Gamma \equiv 1 + \frac{1}{\alpha} > 1$ from Eq.~(\ref{gamma1})
for the IPL potential, and $w_B$ is the equation of state of the fluid
component dominating the background. So, during the era of radiation
domination, with $w_B = \frac{1}{3}$, 
\be
 w_\phi = \frac{\alpha - 6}{3(\alpha + 2)},
\ee
and during the era of matter domination, with, $w_B = 0$, 
\be w_\phi = \frac{-2}{\alpha + 2}\label{track-w}.\ee
We also note here that, as in the case of all tracker potentials, the
tracker solution for the IPL model is approached differently for
different initial conditions. For example, in what is referred to as
the ``overshoot'' case, $\rho_{\phi,I}$ begins from a value greater
than the tracker solution value. Assuming that $\phi$ is released from
rest, the dynamics of the scalar field start with an early kinetic
phase ($\dot{\phi}^2 >> V$) in which $w \rightarrow 1$ so that
$\rho_{\phi} \propto a^{-6}$ (from Eq.~(\ref{rhoi})) and $V$ decreases
very rapidly as $\phi$ runs downhill. Since the kinetic energy is too
large for $\phi$ to join the tracker solution as $\phi$ rolls further
down the potential, $\phi$ will overshoot the  tracker solution. The
field will then freeze (as will $V$ and $\frac{V'}{V}$) as $w_{\phi}$
rushes towards $-1$. Finally, when $\phi$ rejoins the  tracker solution,
$\phi$ will run downhill again and $w_{\phi}$ will increase from -1,
briefly oscillate, and then settle into the tracker value
\cite{steinhardt-1999-59}. 

In the ``undershoot'' case, $\rho_{\phi,I}$
begins from a value much smaller than the tracker solution value, and
$\phi$ is once again  released from rest. This corresponds to the
kinetic energy density being very small and $\phi$, $V$, and
$\frac{V'}{V}$ being approximately constant or ``frozen'' as the
universe evolves. Then, as in the ``overshoot'' case, $w_{\phi}$
reaches close to $-1$, $\rho_{\phi} \approx const.$, and  $\rho_B$ is
decreasing. The value of $w_{\phi}$ then increases from $-1$ as $\phi$ once again
runs downhill. After a few oscillations, $w_{\phi}$ will then rejoin
the tracker solution until $\rho_{\phi}$ becomes the dominant
component in the universe. 

\begin{figure}[h t]
\scalebox{0.21}{\includegraphics{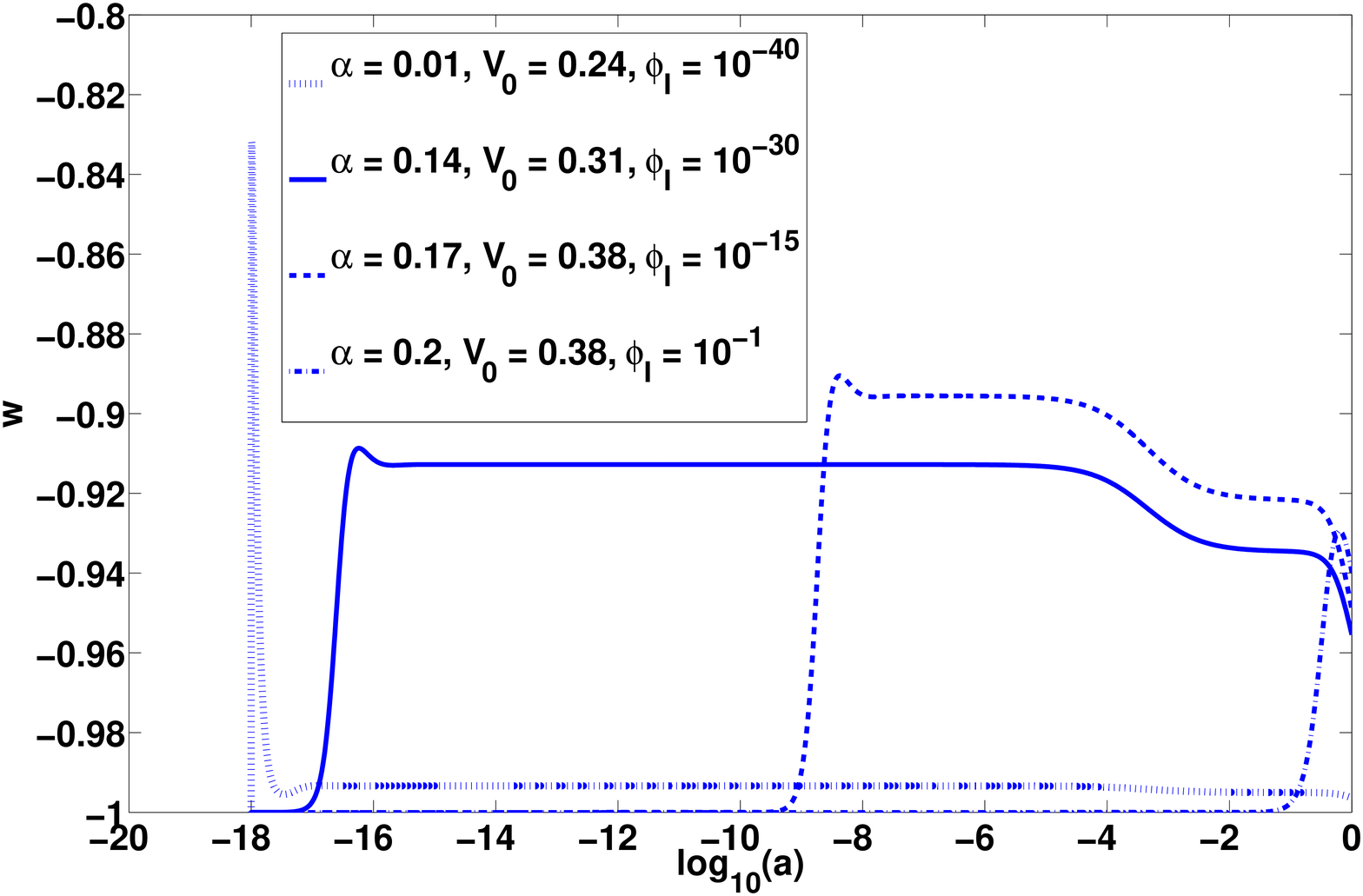}}%
\caption{\label{fig:ipl_fig2} An Illustration of how the evolution and
  tracking behavior of $w$ as a function of scale factor $a$ is
  affected by different  
values of $\alpha$, $V_{0}$, and $\phi_{I}$. The $a$ scale is logarithmic here in order to show behavior on all time scales.}
\end{figure}

\begin{figure}[h t]
\scalebox{0.21}{\includegraphics{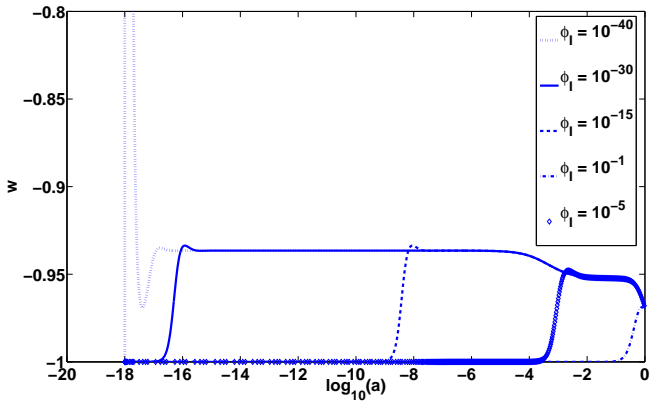}}%
\caption{\label{fig:ipl_fig3} Examples of how the evolution and tracking behavior of $w$ as a function of scale factor $a$ is affected by different values 
of $\phi_{I}$ for given values of $V_{0}$ and $\alpha$. For all
curves, $V_{0} = 0.38$ and $\alpha = 0.1$. These examples illustrate
how different values of  $\phi_{I}$ lead to the same values of the
equation of state parameter today. The $a$ scale is logarithmic here
in order to show behavior on all time  
scales.}
\end{figure}

\begin{figure}[h t]
\scalebox{0.22}{\includegraphics{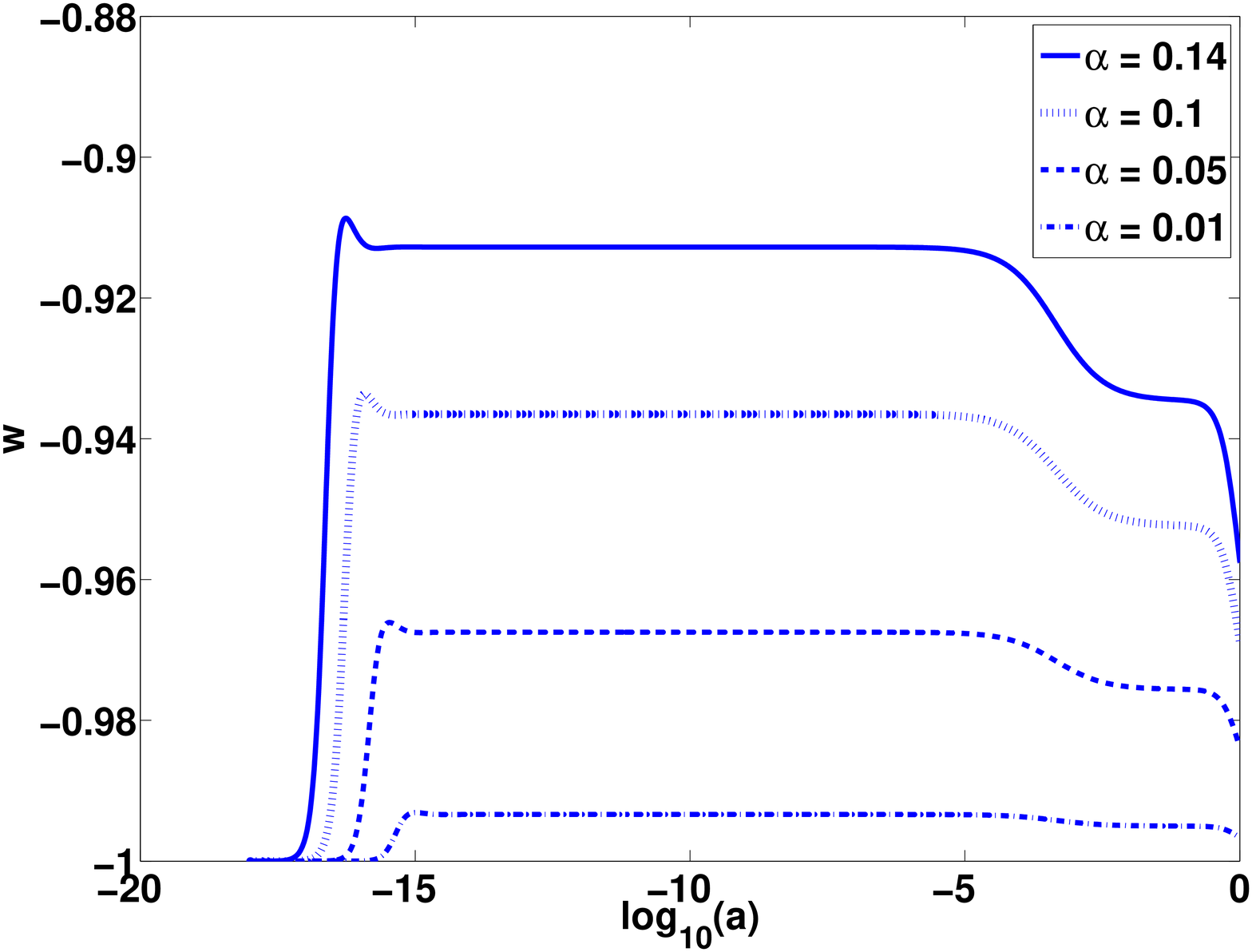}}%
\caption{\label{fig:ipl_fig4} This figure depicts the evolution and
  tracking behavior of $w$ as a function of scale factor $a$ for
  different values of  $\alpha$ for given values of $V_{0} = 0.38$ and
  $\phi_{I} = 10^{-30}$. As long as $\phi_{I} << M_P$, $\alpha$ will
  determine $w_{0}$ and the amplitude  of the $w(a)$ curves. In
  addition, the smaller $\alpha$ is, the later the tracker is reached
  for a given $\phi_{I}$. The $a$ scale is logarithmic here in  
order to show behavior on all time scales.}
\end{figure}

Figures \ref{fig:ipl_fig2},
\ref{fig:ipl_fig3}, and \ref{fig:ipl_fig4} depict the evolution of
$w_{\phi}$ for the IPL model during these various regimes. With 
little sensitivity to the exact value of $V_{0}$, $\alpha$
will determine the amplitude of the $w$ curve and determine the value
of $w_0 \equiv w(z=0) \agt -1$ as long as $\phi_I << M_P$. For given values of
$\alpha$, $\phi_I$ determines when the scalar field joins the tracker
solution and how long it follows the tracking solution
(Fig. \ref{fig:ipl_fig3}). As is pointed out in \cite{malquarti2002},
we also find that for the smaller values of $\alpha$ that we focus 
on in this work (e.g., $\alpha \alt 1$), the smaller $\alpha$ is, 
the later the tracker is reached for a given initial value of $\phi$
(Fig. \ref{fig:ipl_fig4}). With regards to $V_{0}$, we find that while
increasing (decreasing) the value of $V_{0}$ leads to
corresponding increases (decreases) in $\omega_{DE}  = \frac{\rho_{\phi}}{\rho_c}h^2$ at z=0  
(where $h = \frac{H_0}{100}$), as expected,
it leads to very small (essentially negligible) decreases (increases)
in the value of $w_0$ and essentially no change in the tracking
solutions or tracking behavior. When the scalar field has tracking
solutions, different values of $\phi_I$ lead to similar values of, for
example, $-0.9 > w_{0}> -1$, with $w \rightarrow -1$ and $\Omega_{m}
\rightarrow 0$ as $a \rightarrow \infty$. There will be essentially no
dependence of $\phi_I << M_P$ on either the present dark energy
equation of state or the present contribution of dark energy to the
total energy density of the universe (as illustrated in
Fig. \ref{fig:ipl_fig3}).

\subsection{The non-tracking case}

It is possible to find non-tracking cosmological solutions for IPL
quintessence.  If $\phi_I \sim M_P$, then $\phi$ will follow the
tracker solution for only a very brief period of time or 
not exhibit tracking behavior at all.  In our computational
algorithms, for example, we find that tracking solutions do not
strictly exist and thus tracking behavior does not strictly occur for,
roughly, all $\phi_I \agt 10^{-5}$ when $\alpha \alt 1$ and $0.25 \alt
V_{0} \alt 0.45$. Moreover, for some instances in which  $-1.5 \alt
log_{10}(\phi_I) \alt -0.3$, $w \approx -1$ initially but then
increases towards $-1 > w_{0} > -0.9$, for example, as for the case of
``thawing'' models and behaviors
\cite{caldwell2005,scherrer2006}. Examples of this nontracking
``thawing''-like behavior of the equation of state for $\phi_I =
10^{-1}$ for $\alpha = 0.2$ and $0.1$ can also be seen (dashed-dotted
curves) in  Figs. \ref{fig:ipl_fig2} and \ref{fig:ipl_fig3}. 
Nontracking initial conditions for the IPL model as well as possible
connections between the quintessence field and the inflation field
(the inflaton), which is beyond the scope of this work, are discussed
in some detail in \cite{kneller2003} and references therein. Like
\cite{kneller2003}, and as we discuss further in Section
\ref{sec:Sec4a}, we also find that for cases where $\phi_I \rightarrow
M_P$ and the field has not joined the tracker by the present epoch,
the  range of acceptable values of $\alpha$ increases significantly as
$w$ increases. For values of $log_{10}(\phi_I)$ roughly between $-5$
and $-1$, $w_{\phi}$ leaves its tracking phase with matter and enters
a transient phase (see Fig. \ref{fig:ipl_fig3}) before exhibiting
``thawing'' behavior for $log_{10}(\phi_I) \agt -1.5$ 

\subsection{The transition from tracking to acceleration}

For most of this work, we focus on cosmological solutions that exhibit
tracking at early times.  Out of respect for big bang nucleosynthesis
\cite{yahiro2001} and other standard considerations there must be an
early epoch of radiation domination where 
$\rho_{\phi} << \rho_r$ and redshifts as \cite{ratra88,brax1999}
\be \rho_{\phi} \propto a^{-\frac{4 \alpha}{2 + \alpha}}.\ee
It is possible in this case to find an exact solution to the Klein Gordon equation for which 
\be \phi \propto a^{\frac{4}{2 + \alpha}},\ee
and it can be shown that this solution is an attractor \cite{ratra88}.
During matter domination, the attractor is also characterized by the scalar field evolving as
\be \phi \propto a^{\frac{3}{2 + \alpha}},\ee
corresponding to energy density evolving as
\be \rho_\phi \propto a^{-\frac{3 \alpha}{2 + \alpha}}.\ee
As long as $\frac{\rho_\phi}{\rho_m} << 1$, these expressions provide
a very good approximation to the behavior of the IPL quintessence
field  \cite{liddle1998,watson2003}. In other words, the tracking
regime itself is strictly valid only when the expansion of the
universe is dominated by matter.  Then, at later times, when
$\rho_\phi$ starts to make a significant contribution to the cosmic
expansion rate, the  value of $w_\phi$ in Eq.~(\ref{eq-wb}) starts to
diverge from its tracker value, as do $\phi(a)$ and $\rho_{\phi}(a)$,
such that the scalar field mimics  the behavior of a cosmological
constant today (with $w \approx -1$), consistent with current
observations. So, we can see that $\rho_\phi$ in the attractor
solution decreases less quickly than $\rho_m$ and $\rho_r$, which
allows us to realize the following behavior: Deep in the era of
radiation domination, $\rho_\phi$ is small enough to satisfy
constraints from standard models for big bang nucleosynthesis and the
formation of the light elements, but $\rho_\phi$ does eventually
become large enough today (with $w \rightarrow -1$) so that the
universe undergoes accelerated expansion and acts as if it has a
cosmological constant, but one that slowly varies with time and
position \cite{peebles2002}. 

\subsection{Current constraints}

From an observational standpoint, if we require $w_{0}$ to be
roughly consistent with current observational constraints, say, for
example, $-1 \alt w \alt -0.8$, \cite{astier2006,spergel2007,riess2007,kowalski2008} 
then the power law index $\alpha$ must be roughly in the range $0 \alt \alpha \alt 0.5$, yielding a
shallow potential shape. The quintessence equation of state in the current epoch abandons
the tracking regime because the dark energy is now the  dominant
component.  However, the shallow potential
shape makes $w_{0}$ not far from the tracking one in
Eqn.~(\ref{track-w}), differing typically at the 10\% level
\cite{giovi2003,baccigalupi2001}.  

Various combinations of data (including CMB and SNe Ia observations) have
been used to constrain the slope of the IPL potential, finding $\alpha
\alt 1-2$ (e.g., \cite{amendola2003, caldwell2004,colombo2006,calvo2006,schimd2006,schimd2006a,efstathiou2008}), 
so that flatter potentials seem to
be favored by the data. Recently, for example, \cite{colombo2006} have found $0.7
\alt \alpha \alt 0.8$ in an MCMC analysis of the IPL potential when
assuming that the energy scale of the potential is that of a cosmological
constant (i.e.,  $V_0 \approx \Lambda^4 \approx 10^{-47}GeV^4$) and for when
$\Omega_{\phi} = \frac{\rho_{\phi}}{\rho_{c}}$ varies in the range 0.1-0.9 and $h = 0.70$.
A number of authors (e.g., \cite{yahiro2001, bludman2004,
delaMacorra2001}) have argued that such small values of $\alpha$
lead to smaller basins of attraction and thus some degree of fine-tuning
and dependence on initial conditions for the IPL model. We have observed, however, that for the
realistic cosmologies that we consider for this work there remains a
substantial basin of attraction: We can vary the initial conditions over a very large range of
values with the end results for $\Omega_{\phi,0}$,
for example, still being physically acceptable \cite{delaMacorra2001}.

Other authors (e.g.. \cite{steinhardt-1999-59, delaMacorra2001,
zlatev1998}) have also explored a variety of issues
associated with tracking properties and solutions for this model. They considered
theoretical constraints relating to, for example, equipartition initial conditions between
quintessence and the remaining fluid components which argue for larger values of $\alpha$ \cite{steinhardt-1999-59, kneller2003}. 
However, in our work we have focused for the most part on realistic
families of cosmological solutions that are broadly consistent with
observational constraints (i.e., $\alpha \alt 1$) and which also include
IPL tracking properties and behaviors  that give the model its conceptual
appeal. We also note that \cite{caldwell2004} have found that while
$\alpha$ is tightly 
constrained, IPL models with $0.25 \alt \Omega_m \alt 0.4$ remain viable.

The real appeal of IPL models from our point of view is that they
offer an interesting class of non-$\Lambda$ cosmologies with some
degree of theoretical motivation.  Thorough
discussions of the basin of attraction (as well as the still
outstanding cosmological constant problem) are key to a fundamental
understanding of the ultimate importance one might give to
the IPL model.  We regard such discussions as too poorly developed at
this point to give them much weight in the very phenomenological
analysis in this paper.  For our purposes, it is good enough that a
large range if initial conditions can converge to 
a common solution thereby avoiding to a substantial degree the fine tuning of initial values of
$\frac{\rho_{\phi}}{\rho_B}$ and $w_{\phi}$ \cite{kneller2003}.

\section{\label{sec:Sec3}Parameterization of the Inverse Power Law Model}

As a general rule, MCMC analysis requires a careful choice of the
model parameters to be varied. Poor parameter choices and degeneracies
between parameters can slow the rate of convergence and mixing of the
Markov chain, reducing the overall efficiency by which the Markov
chain explores a parameter space. For the IPL  potential, $V = V_0
(\frac{M_P}{\phi})^{\alpha}$, the obvious choice of potential
parameters to be varied is $\phi_I$, $\alpha$, and $V_0$. When we
carried out our MCMC analysis of data forecast by the DETF to
constrain the IPL quintessence model around a fiducial $\Lambda$CDM
model, we chose our fiducial value for $V_0$ (in units of $h^2$) to be
0.38, which is the value of the dark energy density today for a
cosmological constant. We chose to make $V_0$ a model parameter in our
MCMC analysis rather than keeping it fixed because other choices of
$V_0$ could provide equivalent cosmological solutions, and we were
also interested in ascertaining how the MCMC exploration of the
parameter space and its ability to constrain the other parameters
would be affected by varying $V_{0}$ as well.  

We have not found a need to reparameterize the IPL 
parameters to the extent that has
been done, for example, in \cite{bozek2007,barnard2007} for the
Albrecht-Skordis or Exponential potential quintessence models. We did,
however, find it necessary to place bounds on some of the potential parameters
in order to prevent the MCMC from infinitely stepping into divergent
directions of parameter space and thus never converging to a
stationary probability distribution. Another reason we placed bounds
on the potential parameters was to prevent the MCMC from spending
possibly large amounts of computer time exploring uninteresting
regions of parameter space that may be completely inconsistent with
observational and theoretical constraints. 

We placed a lower bound of $0$ on $\alpha$, as $\alpha > 0$ is
required for the pure IPL model that we consider
\cite{peebles87}. Given that the DETF data used in the first part of
our MCMC analysis is modeled around a cosmological constant, the most
probable values of $\alpha$ will be those in which $\alpha$ approaches
zero. From Eq.~(\ref{vipl}) we see that as $\phi_{I} \rightarrow M_P$
any value of $\alpha$ will lead to the same value of the 
potential $V(\phi)$ for a given $V_{0}$. However, since $\alpha$ largely
controls the shape of the potential (as well as
the amplitude of $w(a)$) and thus the evolution of the dark energy
density and $w_{\phi,0}$, we find that the simulated data sets place
sufficient constraints on $\alpha$ to prevent the MCMC from infinitely
stepping into divergent directions in the $\alpha - \phi_I$ and $V_{0}
- \phi_I$ parameter spaces even when $\phi_{I} \rightarrow M_P$. This
renders a stringent upper bound  on $\alpha$ 
unnecessary. 

We can also see from Eq.~(\ref{vipl}) that $\phi_I$ can
take on any value and lead to solutions indistinguishable from a
cosmological  constant as $\alpha \rightarrow 0$. This degeneracy
leads to a divergent direction in the $\alpha - \phi_I$ space, where
$\phi_I$ can be arbitrarily large or small. Also,  the simulated data
sets do not constrain $\phi_I$ nearly as tightly as $\alpha$ due to
the fact (previously discussed in the context of attractor
solutions) that a broad  range of $\phi_I$ values can lead to the same
$\phi_{0}$ and $w_{0}$ and thus have little effect on the evolution
of the dark energy density.   Because of this effect, it is necessary   to choose some
cut-offs on $\phi_I$ so that these infinite directions are bounded. 

As discussed in Section \ref{sec:Sec2b} we have parameterized our
potential in a way that gives cosmologically
realistic solutions where $V(\phi)$ approaches the value of the dark energy density
today when $\phi \approx M_P$.  With this in mind, we impose an upper
bound of $M_P$ on $\phi_I$ which helps avoid solutions with
uninterestingly low values of $\rho_\phi$ as well as solutions that are
dominated by transients.
We also note here that, given that the main thrust of our work involved an
MCMC analysis of the regions of parameter space associated with
tracking, we have selected or filtered out  non-tracking parameter
values in the algorithms used to generate likelihood contours from the
MCMC chains by implementing in our algorithms the criteria for tracking 
solutions (as discussed in Section \ref{sec:Sec2a} and \ref{sec:Sec2b}) and, 
specifically, the ``equation of motion'' discussed in \cite{steinhardt-1999-59}. 
Thus, all of the error contours displayed and discussed
in sections \ref{sec:Sec4a} and \ref{sec:Sec4b} correspond to portions
of the parameter space associated with tracking (i.e., parameter
values corresponding  to attractor solutions of the Klein-Gordon
equation). Incidentally, we have found that for a typical Stage 2 MCMC
chain generated from a $\Lambda CDM$ model, for example,  about $90\%$
of points stepped to in the chain correspond to parameters with tracker
solutions, whereas the other $10\%$ correspond to non-tracking
(transient and thawing) parameters. 

Regarding a lower bound on $\phi_I$, we recall from
Section \ref{sec:Sec2b} that we must have $\phi_I << M_P$ so that
the present field value, $\phi_0$ (of order $M_P$), is reached
from a very broad range of initial conditions. This insures that the
tracking properties and solutions that make this model appealing are
still included and valid within the parameter space explored in our
MCMC analysis. If the lower bound on
$\phi_I$ is too large $\phi_I$ may reach the tracking phase only at
very late times or only by the present time (or not at all), leading
to a small basin of attraction and fine-tuning problems. We find that
placing a lower bound of $\phi_I 
= 10^{-20}$ in our MCMC analysis gives reasonable results by ensuring
that on the one hand the tracking solutions and properties are
included in the parameter space explored by the MCMC (i.e., there is a
larger basin of attraction and $\phi_0 \approx M_P$) but on the
other hand, an appropriate cut-off or bound has been placed on a
divergent direction in the $\alpha - \phi_I$ space that may not
otherwise be constrained by the data (and thus possibly preventing the
MCMC chains from coming to equilibrium). 

The above lower bound is not well suited for examining the
finer details of nontracking transient and ``thawing'' regions of
parameter space (where $\phi_I \rightarrow M_P$). 
In chains with a lower bound of $10^{-20}$ or smaller  on $\phi_I$,
the part ($\approx 10\%$) of the chain that shows nontracking and thawing
IPL solutions is not sufficiently well populated to show the full
structure of the probability distribution.   In order to allow the
MCMC to step more frequently in these parameter space regions and so
better converge (as
discussed in \cite{abrahamse2007}) on a well-resolved probability distribution  for the
nontracking and thawing regions of the parameter space, we have also
carried out an MCMC analysis with a lower bound of $-3$ placed on
$log_{10}(\phi_I)$ (see Section \ref{sec:Sec4c}). 

\section{\label{sec:Sec4}MCMC Results and Analysis}

\subsection{\label{sec:Sec4p1}General approach}

Following the approach taken by the DETF, we generated ``data models''
or simulated data sets for future SNe Ia, baryon acoustic oscillation (BAO),
weak gravitational lensing (WL), and CMB (PLANCK) observations. These
considerations of DE projects follow developments in ``stages'': Stage
2 represents ongoing projects that are relevant to dark energy; Stage 3 
consists of medium-cost, near-term, currently proposed
projects (such as BAO,  SNe Ia, and WL surveys with 4-meter class
telescopes using photometric redshifts); Stage 4 consists of a Joint
Dark Energy (Space) Mission (JDEM),  Square Kilometer Array (SKA),
and/or Large Survey Telescope (LST) \cite{DETF06}. ``Optimistic'' and
``pessimistic'' versions of the same data models give different
estimates of systematic errors. Additional information on the
specific DETF data models is given in Appendix A of
\cite{abrahamse2007} and the technical appendix of the DETF report
\cite{DETF06}. We excluded the DETF galaxy cluster data models in our
work because the extension of the DETF calculations to our analysis is
not straightforward, especially in regards to estimates of systematic
errors \cite{albrec2007,abrahamse2007,bozek2007,barnard2007}.  

We have generated two sets of data models. One type is generated
around a cosmology with a cosmological constant, consistent with DETF
Stage 2, 3, and 4 SNe Ia, WL, BAO, and CMB data models. The other set
of data models is built around an IPL fiducial model which was chosen
to be consistent with simulated Stage 2 data based on a cosmological
constant cosmology. We
then use an MCMC algorithm to map the likelihood around each fiducial
model ($\Lambda CDM$ and IPL) via a Markov chain of points in
parameter space, starting with the fiducial model and moving to a
succession of random points in space using a  Metropolis-Hastings
stepping algorithm. The technical details of our MCMC algorithm are
presented in Appendix B of \cite{abrahamse2007} and references
therein. In this way we can, for example, analyze the parameter space
of IPL quintessence in the light of DETF data models and evaluate the
likelihood function of the parameters of our model. Once the Markov
chains of our models in parameter space have been
computed we can extract likelihood contours from the
distribution of models and display them as projected 2-D likelihood contour
plots. This can then give us a picture of the shape of the likelihood
region of all the parameters in our models in the whole
multidimensional parameter space if we were to plot likelihood
contours for each pair of parameters in the parameter space. In all
plots in this paper, we show $68.27\%$ $(1\sigma)$, $95.44\%$
$(2\sigma)$, and $99.73\%$ $(3\sigma)$ confidence contours, which
consist of points where the likelihood equals $e^{-\frac{2.30}{2}}$, 
$e^{-\frac{6.17}{2}}$, and  $e^{-\frac{11.8}{2}}$ of the maximum value
of the likelihood, respectively. We have constructed these plots by
marginalizing over all of the  cosmological parameters, $\omega_m$,
$\omega_k$, $\omega_B$, $\omega_r$, $h$, $\delta_\zeta$, $n_s$,
$n'_s$, (as defined by the DETF), and the various nuisance and/or
photometric redshift parameters, which take into account uncertainties
and errors in the simulated data. The nuisance and photometric
redshift parameters are described and explained in detail in
\cite{abrahamse2007,DETF06}.  
  
\subsection{\label{sec:Sec4a}Cosmological Constant Fiducial Model}

In this section we present the results of our MCMC analysis for the
combined simulated data sets generated around a $\Lambda$CDM
cosmology. We list the values of the free parameters for our
$\Lambda$CDM fiducial model (with energy density and $V_{0}$ in units of $h^2$
and $\phi_I$ in reduced Planck units) in Table \ref{lamfidiplpars}.  
(The IPL parameters given generate a cosmological constant.)

\begin{table}[ht]
\centering
\caption{Fiducial Parameter Values (energy densities in units of $h^2$) for $\Lambda CDM$ model.}
\begin{tabular}{|l l|}
        \hline \hline
$\omega_{DE}$ & $0.3796$              \\
$\omega_{m}$ & $0.146$                \\ 
$\omega_{k}$ & $0.0$                  \\
$\omega_{B}$ & $0.024$                \\
$\omega_{r}$ & $4.16 \times 10^{-5}$  \\
$n_s$        & $1.0 $                 \\
$n'_s$       & $0.00001$              \\
$\delta_{\xi}$  & $0.87$              \\
$h$ & $0.72$                          \\
$\alpha$     & $0.0$                  \\
$\phi_I$     & $10^{-15}$             \\ 
$V_{0}$      & $0.38$                 \\
        \hline
\end{tabular}
\label{lamfidiplpars}
\end{table}

We note that $h^{2}(a=1) = \omega_{m} + \omega_{r} + \omega_{k} +
\omega_{DE}$, with recent observations providing a prior constraint of
$h = 0.72 \pm 0.008$ \cite{Freedman2000}. Also, $\omega_{r}$, the
radiation energy density, is not a free parameter for our calculations
but is fixed by the CMB temperature (and the standard assumption of
three massless neutrinos) \cite{DETF06}.  

Stage 2 combines SNe Ia, WL, and CMB data models but does not include
BAO data models. Stages 3 and 4 additionally include the BAO data
models as well. As discussed in Section \ref{sec:Sec4p1} we project
our probability distributions into 2-D spaces given by pairs of the
IPL parameters (i.e., the $V_{0}-\alpha$,
$V_{0}-\phi_{I}$, and $\phi_{I}-\alpha$ planes).
  
\begin{figure}[h t]
\scalebox{.28}{\includegraphics{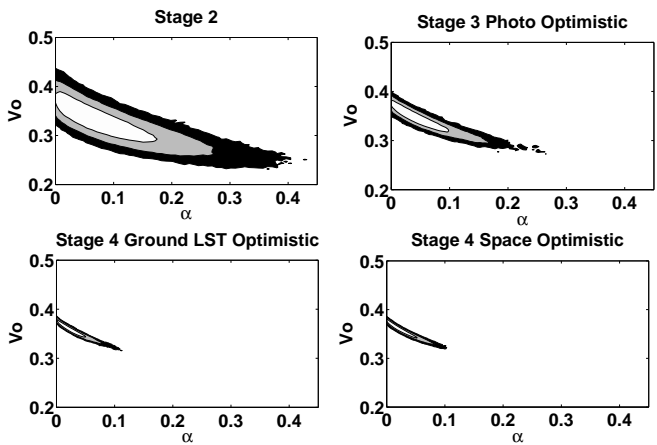}}%
\caption{$V_{0}-\alpha$ $1 \sigma (68.27\%)$, $2 \sigma (95.44\%)$, and $3 \sigma (99.73\%)$ confidence regions
  for DETF ``optimistic'' combined $\Lambda CDM$ data models. \label{V0_alpha_opt_lambdafid_cut4}}
\end{figure}

The likelihood contours in the $V_{0}-\alpha$ plane, with all
non-tracking parameter values ($log_{10}(\phi) \alt -6$) excluded, for
Stage 2 and the optimistic versions of Stage 3 photometric, Stage 4
Space, and Stage 4 Ground LST combined data are shown in
Fig. \ref{V0_alpha_opt_lambdafid_cut4}. In all cases the error contours
show the expected trend of the IPL potential to approach a cosmological constant as
$\alpha \rightarrow 0$ (and also corresponding to where the slope of
the potential goes to $0$). The vertical axis where $\alpha = 0$
corresponds to $\Lambda$. Therefore, the value of $V(\phi_0) = V_{0}$
on the vertical axis represents $\Lambda$ or the dark energy density
$\omega_{DE}$ for $\alpha = 0$. However, along the lines of the
discussion in Section III of \cite{barnard2007} for the
Albrecht-Skordis model and as discussed in this paper in Section
\ref{sec:Sec2b}, we must also keep in mind that the parameter $V_{0}$
does not have a significant effect on the equation of state of dark
energy. Moreover, for $\alpha > 0$, $V_{0}$ is no longer identical to
$\omega_{DE,0} \equiv \omega_{DE}(z=0)$.  

For small values of $\alpha$, there is a spread in $V_{0}$ in the $V_{0}-\alpha$ space. Since
these values of $\alpha$ are consistent with $\Lambda$ or a
non-evolving dark energy, the spread in $V_{0}$ is essentially a
measure of how well the experiments are measuring $\omega_{DE,0}$. The
spread or uncertainty in $V_{0}$ for all $\alpha$ is also a result
of uncertainties on  measurements of $\Omega_{m,0}$. Larger values of
$\alpha$ correspond to larger values of $w$ ($w > -1$) and thus values
of $w$ that deviate more and more from the equation of state for
$\Lambda$ as $\alpha$ increases, possibly up to values of $\alpha$
that correspond to detectable differences from $\Lambda$. The smallest
values of $V_{0}$ correspond to the largest values of $\alpha$, which
in turn correspond to the largest values of the equation of state (and
hence those values of $w$ deviating the most from what we expect for a
cosmological constant). As the value of $\alpha$ increases, we see
that the likelihood contour in Fig. \ref{V0_alpha_opt_lambdafid_cut4}
has an overall downward curved shape. This is due to the fact
that the slope of the potential becomes steeper for increasing values
of $\alpha$, which leads to greater evolution of the dark energy
density and larger values of $w_0$ that deviate more and more from $-1$.
The reduction in the  $V_{0}$ direction reflects improving constraints
with increasing stage number on the dark energy density. As a specific
quantitative example of this, we see from the Stage 4 error contours in
Fig. \ref{V0_alpha_opt_lambdafid_cut4} that the extrema of the range
of $V_0$ values deviates from the fiducial value by less than $20\%$
when $\alpha \approx 0.1$ and less than $5\%$ when $\alpha \approx 0$.   
The shrinking in the $\alpha$ direction corresponds to increasing constraints on
deviations from a cosmological constant.   

\begin{figure}[h t]
\scalebox{.26}{\includegraphics{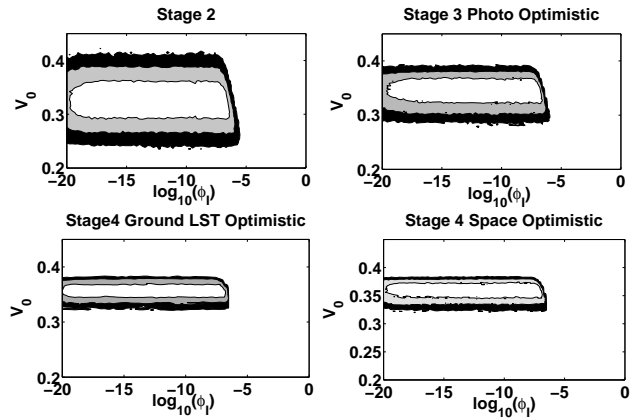}}%
\caption{$V_{0}-log_{10}(\phi_{I})$ $1 \sigma (68.27\%)$, $2 \sigma (95.44\%)$ and $3 \sigma (99.73\%)$ confidence regions
  for DETF ``optimistic'' combined $\Lambda CDM$ data models. \label{V0_phiI_opt_lambdafid_cut4}}
\end{figure}

Fig. \ref{V0_phiI_opt_lambdafid_cut4} depicts likelihood contours in
$V_{0}-log_{10}(\phi_{I})$ space, where, again, all non-tracking
transient and thawing parameter values have been removed. As
noted in Section \ref{sec:Sec3},  we imposed $10^{-20}<
\phi_I/M_P < 1$. Since a large range of initial values of the  scalar field
($\phi_{I} < M_P$) are generally washed out by the tracking behavior,
we can see from the contours that 
there is  very little dependence of the dark energy density today
on $\phi_I << M_P$. Once again the spread in $V_{0}$ values is
essentially a measure of how well the experiments are measuring the
dark energy density at the present time. The error contours also show
a slight trend toward an increasing range of acceptable values of
$\phi_I$ which possess attractor solutions as $V_0$ decreases, which
is associated with greater $\alpha$ values and thus greater dark
energy evolution.  The sections of the overall parameter space depicted
in these figures also tend to disfavor larger values of $\alpha$, or,
equivalently, disfavor larger departures from a cosmological constant
and thus more dark energy evolution. We once again note a reduction in
the $V_{0}$ direction with increasing stage number, indicating the
improving constraints that the data places on the dark energy
contribution to the total energy density of the universe today.  

\begin{figure}[h t]
\scalebox{.26}{\includegraphics{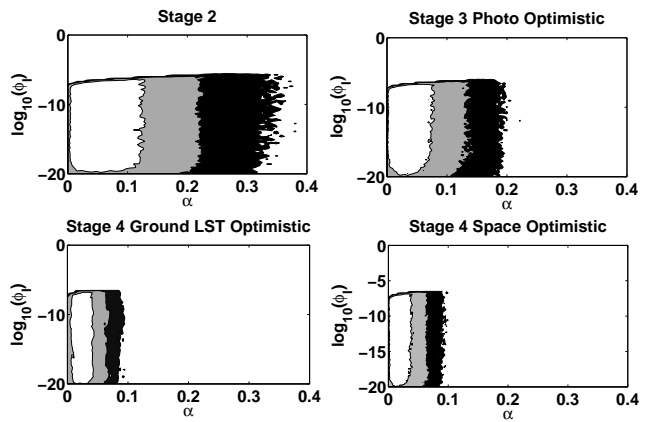}}%
\caption{$log_{10}(\phi_{I})-\alpha$ $1 \sigma (68.27\%)$, $2 \sigma (95.44\%)$ and $3 \sigma (99.73\%)$ confidence regions
  for DETF ``optimistic'' combined $\Lambda CDM$ data models. \label{phiI_alpha_opt_lambdafid_cut4}}
\end{figure}


The likelihood contours in the $log(\phi_{I})-\alpha$
(Fig. \ref{phiI_alpha_opt_lambdafid_cut4}) space are clearly seen to
shrink in the $\alpha$ direction with increasing stage number, once
again showing improving constraints on the amount of dark energy
evolution and on deviations from a cosmological constant from Stage 2
to Stage 3 and from Stage 3 to Stage 4. This corresponds to a greater
disfavoring of larger values of $\alpha$ with successive stages of
data. We also see in  Fig. \ref{phiI_alpha_opt_lambdafid_cut4} a very
slight trend toward an increasing range of acceptable values of
$\phi_I$ possessing attractor solutions as $\alpha$ increases. This
corresponds to the trend of a larger range and upper limit for
$\phi_I$ having attractor solutions for smaller values of $V_0$
discussed in regards to  Fig. \ref{V0_phiI_opt_lambdafid_cut4}. 

The
trend in Fig. \ref{phiI_alpha_opt_lambdafid_cut4} is related to the
fact that the largest values of of $\phi_I$ from which the attractor
is joined before the present time occur on the flatter portions of the
potential where $w_{\phi}$ is closer to $-1$ in recent times and today
and the curvature and slope of the potential is smaller. Since
$\alpha$ controls the steepness of the potential, changes in $\alpha$
have less of an effect on the flatter parts of the potential where
$\phi_I$ is larger and $V(\phi_I)$, $V(\phi_I)'$, and $V(\phi_I)''$
are smaller (as can also be seen in
Eq.~(\ref{vtrack})-Eq.~(\ref{curvature})). So, when the scalar field
tracks the background evolution on flatter portions of the potential,
we expect a slight  increase in the range of acceptable (tracking)
$\phi_I$ values as $\alpha$ increases.    

Overall, as found by the DETF, successive stages of data do better at constraining the
evolution of dark energy. As can be seen in the likelihood contours
above and as was also found for the case of the Albrecht-Skordis model
\cite{barnard2007}, the IPL potential parameters appear to be somewhat
better constrained by the DETF Stage 4 LST ground data models than by
the DETF Stage 4 space data models. This reflects the fact that ground
and space data are sensitive to slightly  different features of the
dark energy evolution.

\subsection{\label{sec:Sec4b}Inverse Power Law Fiducial Model}

We next evaluate the power of future experiments by assuming that the dark energy in the universe can actually be described by the inverse power law model
rather than a $\Lambda CDM$ fiducial model. For our fiducial IPL model, we use $\alpha = 0.14$, $\phi_I = 10^{-15}$, and $V_{0} = 0.31$. The remaining parameters
of the IPL fiducial model are the same as those used in the fiducial $\Lambda CDM$ model. Our IPL model fiducial values (given in Table \ref{iplfidpars} with energy
densities and $V_{0}$ in units of $h^2$ and $\phi_{I}$ in reduced Planck units) were chosen, excluding consideration of the ``thawing''
or outlying regions of the parameter space, to lie near the boundary of (or just beyond) 1 $\sigma$ detection or
within the $95.44\% (2\sigma)$ confidence region in the $V_0-\alpha$ and $log(\phi_I)-\alpha$ spaces (Fig. \ref{V0_alpha_opt_lambdafid_cut4} and 
Fig. \ref{phiI_alpha_opt_lambdafid_cut4}) for Stage 2 $\Lambda CDM$ data, but excluded by more than $3 \sigma$ in the Stage 4 optimistic ground and space data so as to 
be strongly ruled out by Stage 4 $\Lambda CDM$ data.

\begin{table}[ht]
\centering
\caption{Fiducial Parameter Values (energy densities in units of $h^2$) for Inverse Power Law model.}
\begin{tabular}{|l l|}
        \hline \hline
$\omega_{DE}$ & $0.3796$              \\
$\omega_{m}$ & $0.146$                \\ 
$\omega_{k}$ & $0.0$                  \\
$\omega_{B}$ & $0.024$                \\
$\omega_{r}$ & $4.16 \times 10^{-5}$  \\
$n_s$        & $1.0 $                 \\
$n'_s$       & $0.00001$              \\
$\delta_{\xi}$  & $0.87$              \\
$h$ & $0.72$                          \\
$\alpha$     & $0.14$                 \\
$\phi_I$     & $10^{-15}$             \\ 
$V_{0}$      & $0.31$                 \\
        \hline
\end{tabular}
\label{iplfidpars}
\end{table} 

\begin{figure}[h !]
\centerline{ 
\scalebox{0.21}{\includegraphics{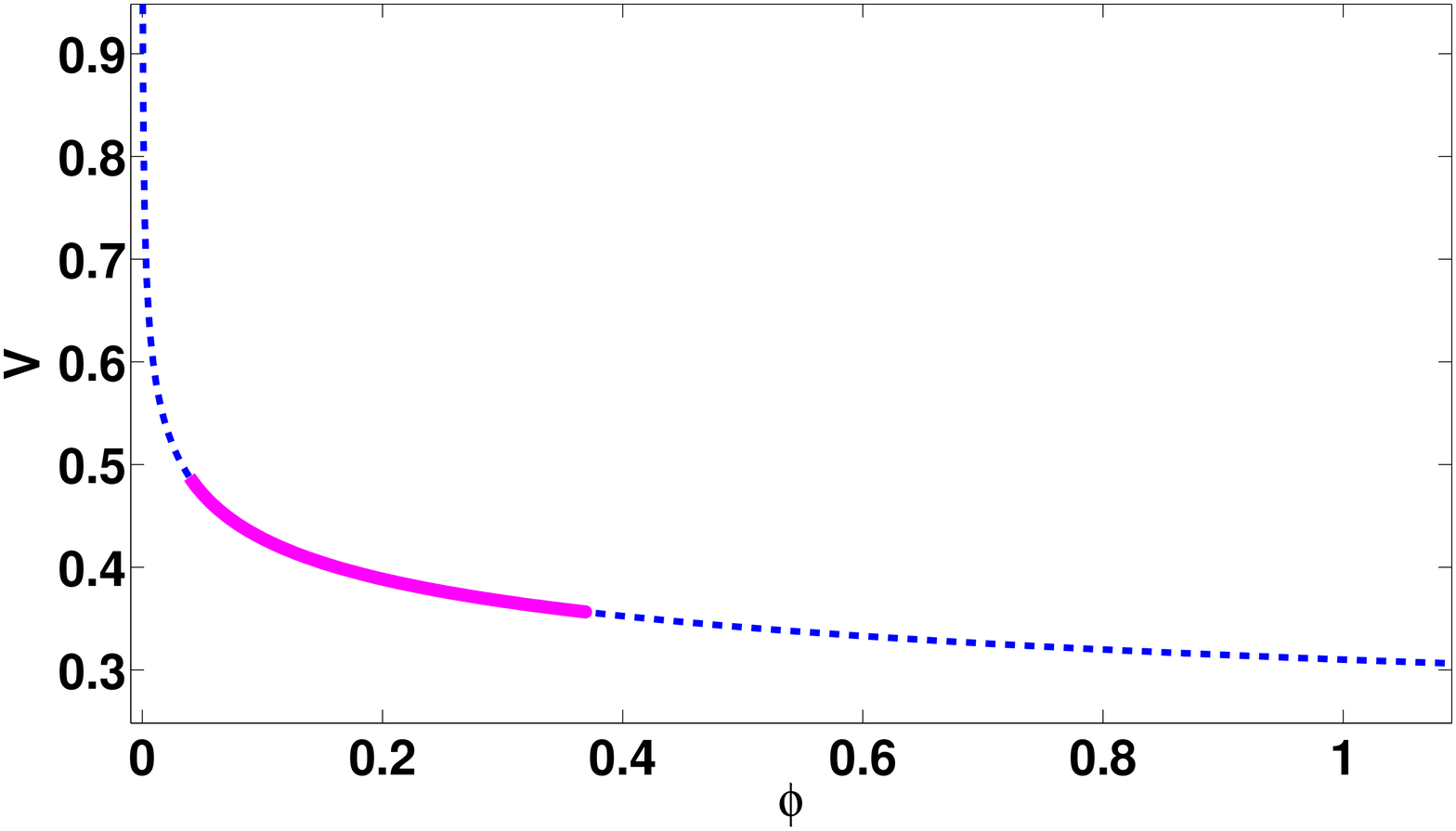}}
}
\hfill
\vfill
\centerline{
\scalebox{0.21}{\includegraphics{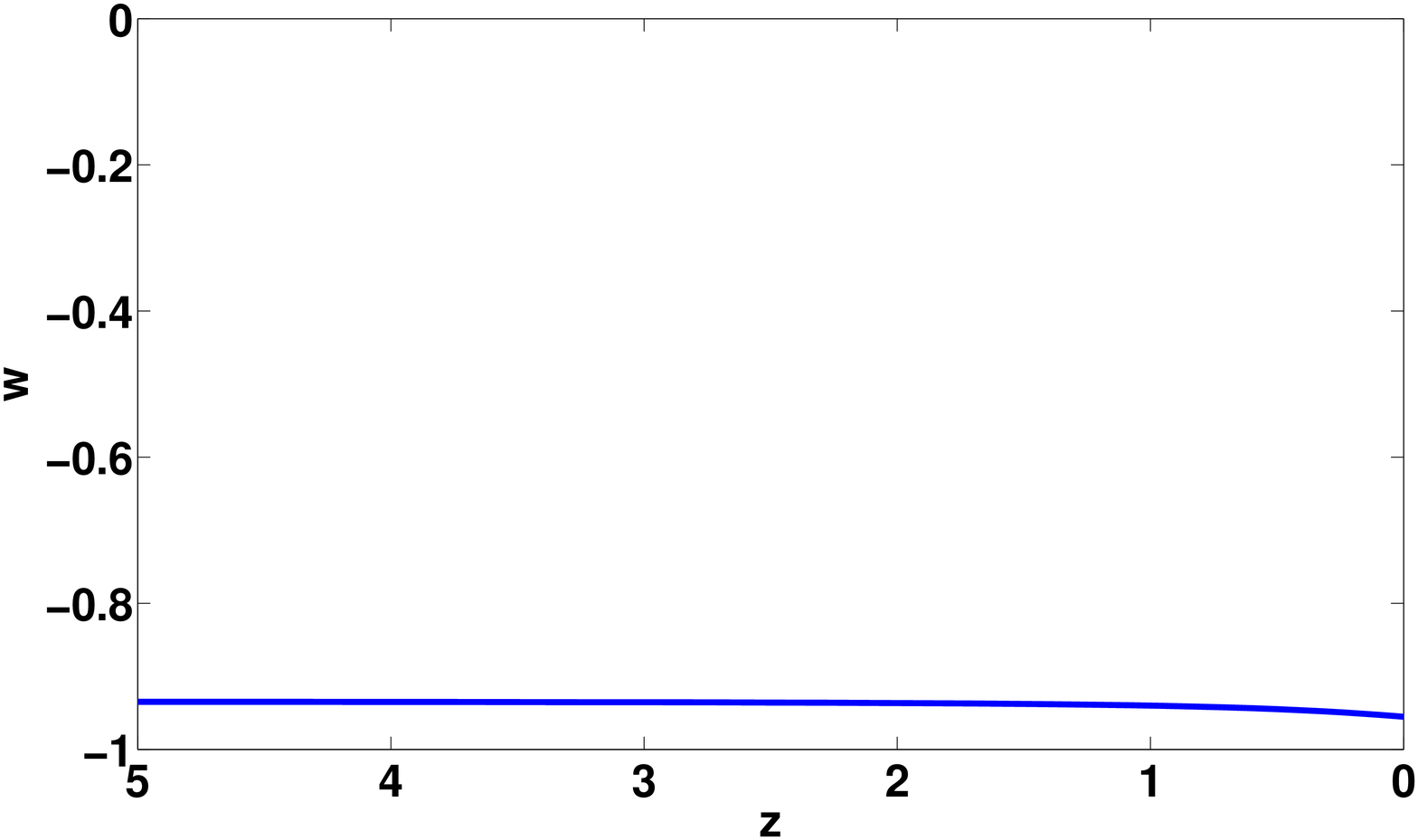}}
}
\caption{\label{fig:fid_ipl_Vw} The potential of the IPL fiducial
  model ($\alpha = 0.14$, $\phi_I = 10^{-15}$, $V_{0} = 0.31$) (top
  panel,dashed  curve). The corresponding equation of state evolution
  $w(z)$ for a potentially observable range of redshift values is
  shown in the bottom panel.  The solid curve overlaying the potential
  in the top panel shows the evolution of the IPL fiducial model
  scalar field for the range of $z$ values  (from $z = 5$ to the
  present time) depicted for $w(z)$ in the bottom panel.} \end{figure}

We also ensured that this fiducial model had initial conditions and
had an equation of state such that the attractor is joined before the
present time. The equation  of state parameter as a function of scale
factor $a$ for all time scales (the $a$ scale is logarithmic) for our
fiducial model is similar to the dashed curves in
Figs. \ref{fig:ipl_fig2} and \ref{fig:ipl_fig3}. We also depict the
potential of the fiducial model in the top panel of
Fig. \ref{fig:fid_ipl_Vw} along with the corresponding equation of
state evolution as a function of redshift in the bottom panel. The
fiducial model corresponds to the point  $w_{0} = -0.955$, which deviates from 
$w(z) = -1$ by only about $4.5\%$. We have chosen our fiducial model to thus be marginally consistent with
the $\Lambda CDM$-based data but demonstrating enough dark energy
evolution to be different enough from $\Lambda$ to be resolved by
Stage 4 experiments. In this way we are able to illustrate the power
of Stage 4 data models and their ability to rule out the $\Lambda$ model.

\begin{figure}[h !]
\scalebox{0.27}{\includegraphics{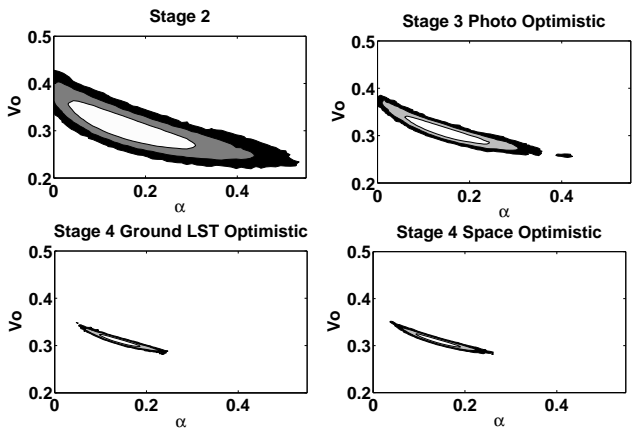}}%
\caption{$V_{0}-\alpha$ $1 \sigma$ $(68.27\%)$, $2 \sigma$ $(95.44\%)$ and $3 \sigma$ $(99.73\%)$ likelihood contours
  for DETF optimistic combined data sets generated from a selected IPL background cosmological model. \label{V0_alpha_opt_iplfid_cut4}}
\end{figure}

  Duplicating our MCMC analysis methods for the IPL fiducial model, we
  again marginalized over all but two pairs of the parameters
  $\alpha$, $\phi_{I}$, and $V_{0}$ for the purposes of generating 2-D
  likelihood regions for the IPL dark energy
  parameters. Fig. \ref{V0_alpha_opt_iplfid_cut4} shows the results of
  our MCMC  analysis and calculations for Stage 2, Stage 3
  Photo-optimistic, Stage 4 LST Optimistic, and Stage 4 Space
  Optimistic data models in the $V_{0}-\alpha$ parameter  space. We
  can see from the $\alpha = 0$ axis, corresponding to a cosmological
  constant, that the $\Lambda CDM$ model (i.e., a non-evolving scalar
  field) is still allowed at Stage 2 (at the $2\sigma$ $(95.44\%)$
  confidence level but not quite at the $1\sigma$ $(68.27\%)$
  confidence level) but  becomes less favored by subsequent stages of
  data models. At Stage 3 the $\Lambda CDM$ model lies outside of the $2
  \sigma$ contour, and by Stage 4 it is ruled out by well over $3
  \sigma$. For Stage 2 and subsequent stages the range of $\alpha$
  values covered by  the contours is significantly greater than for
  the $\Lambda CDM$ case since dark energy solutions with more
  evolution are favored more here. The greater dark energy  evolution
  for this case also leads to the slightly more significant downward
  trend in the shape of the contours than is seen in the $\Lambda CDM$
  confidence contours.  

The described increase in constraining power for higher quality data models is
similar to the $\Lambda CDM$ results in Section \ref{sec:Sec4a}
for the  $\Lambda CDM$ model. However, as previously indicated, the
range of $\alpha$ values has significantly increased within the $1
\sigma$,  $2 \sigma$, and $3 \sigma$ contours, allowing for an
increased range of evolving dark energy solutions. By the Stage 4
combined data sets, we can clearly differentiate between our selected
IPL fiducial model and the $\Lambda CDM$ model by well over $3
\sigma$. This increased constraining power is again consistent with
the ($\Lambda CDM$) DETF results for Stage 4 experiments.  Hence, the
results of our MCMC analysis, as seen in
Fig. \ref{V0_alpha_opt_iplfid_cut4} (as well as
Fig. \ref{phiI_alpha_opt_iplfid_cut4} below), show that, \textit{for a
universe described by this specific IPL fiducial model, the Stage 4
experiments will rule out a cosmological constant by well over $3
\sigma$}.

\begin{figure}[h !]
\scalebox{0.26}{\includegraphics{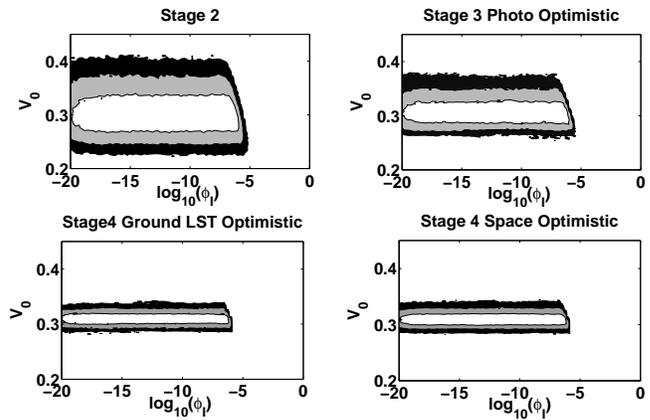}}%
\caption{$V_{0}-log(\phi_I)$ $1 \sigma$ $(68.27\%)$, $2 \sigma$ $(95.44\%)$ and $3 \sigma$ $(99.73\%)$ likelihood contours
  for DETF optimistic combined data sets generated from a selected IPL background cosmological model. \label{V0_phiI_opt_iplfid_cut4}}
\end{figure}

Figure \ref{V0_phiI_opt_iplfid_cut4} shows likelihood contours in
$V_{0}-log(\phi_I)$ space. As for the case of the $\Lambda CDM$ model,
there is again very  little dependence of dark energy density
today on $\phi_I$ when $\phi_I << M_P$. Once again, the spread in
$V_{0}$ values is essentially a measure of how well the  experiments
are measuring the present dark energy density as given by the chosen
IPL fiducial model.  The trend toward an increasing
range of and acceptable upper limit to values of $\phi_I$ possessing
attractor solutions for smaller $V_0$, as noted in reference to
Fig. \ref{V0_phiI_opt_lambdafid_cut4}, is slightly more pronounced here due
to the larger range of acceptable values of $\alpha$ and greater DE
evolution for the IPL model.   

The likelihood contours in the $log(\phi_I)-\alpha$
(Fig. \ref{phiI_alpha_opt_iplfid_cut4}) plots are clearly seen to
shrink in the $\alpha$ direction with increasing stage number, but the
overall range of $\alpha$ values stepped to by the MCMC chain and thus
included within the likelihood contours is significantly larger than
for the $\Lambda CDM$ model data sets, again indicating that dark
energy solutions with more evolution are disfavored less for this IPL
fiducial model than for the $\Lambda CDM$ model. The
$log(\phi_I)-\alpha$ contours also show (like the $V_{0}-log(\phi_I)$
contours in Fig. \ref{V0_phiI_opt_iplfid_cut4}) that the $\Lambda CDM$
model is still allowed at Stage 2 (but lies just outside of the $1 \sigma$
contour) and at Stage 3 Photo-optimistic (lying outside of the $2
\sigma$ contour here) but is ruled out by well over $3 \sigma$ by
Stage 4, again becoming less favored by subsequent stages of data
sets. We also see a slightly more pronounced trend of an increasing
range of acceptable values of $\phi_I$ possessing attractor solutions
as $\alpha$ increases. Again, this corresponds to the increasing range
of acceptable $\phi_I$ values possessing attractor solutions for
smaller $V_0$ in Fig. \ref{V0_phiI_opt_iplfid_cut4} and the fact that the
part of the IPL potential where the largest values of $\phi_I$ that
lead to attractor solutions that are still acceptable is steeper
(larger $\alpha$) than for the $\Lambda CDM$ case
(Eqn.~(\ref{vtrack})-Eqn.~(\ref{curvature})).  

\begin{figure}[h !]
\scalebox{0.26}{\includegraphics{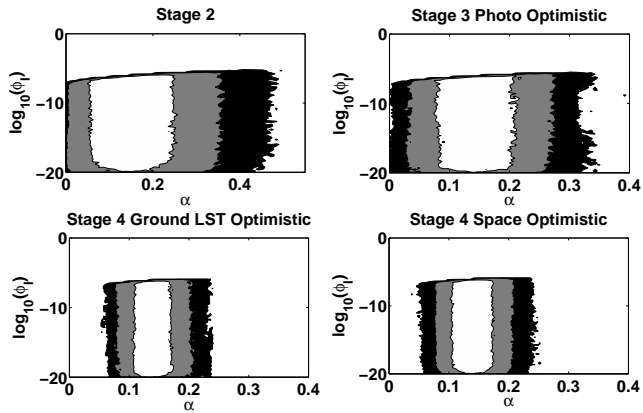}}%
\caption{$log(\phi_I)-\alpha$ $1 \sigma$ $(68.27\%)$, $2 \sigma$ $(95.44\%)$ and $3 \sigma$ $(99.73\%)$ likelihood contours
  for DETF optimistic combined data sets generated from a selected IPL background cosmological model. \label{phiI_alpha_opt_iplfid_cut4}}
\end{figure}


As in the case of the $\Lambda CDM$ data sets discussed in Section
\ref{sec:Sec4a} and was found in the Albrecht-Skordis model
\cite{barnard2007}, we find once again that Stage 4 ground data (this
time based on our fiducial IPL model) slightly more strongly
constrains the parameters $\alpha$ and $V_{0}$ than does the Stage 4
space data. This is opposite of what has been found with other scalar
field models \cite{abrahamse2007,bozek2007}.

\subsection{\label{sec:Sec4c}Non-Tracking Parameter Space Regions}

Though the main focus of our work has involved an analysis of the
tracking regions of the parameter space of the IPL model, here we
discuss briefly the  results of our MCMC analysis of the non-tracking
regions, i.e., initial values of the scalar field from which the
attractor is \textit{not} joined before or by the present time.  In
this case, our motivation is simply to explore an interesting-looking
class of dark energy behaviors that have already been considered
elsewhere in the literature (e.g., \cite{kneller2003}). We
acknowledge that to the extent that the tracking 
behavior is a key reason to consider the IPL model, the solutions
considered in this section do not benefit from the same degree of motivation.

\begin{figure}[h !]
\scalebox{0.27}{\includegraphics{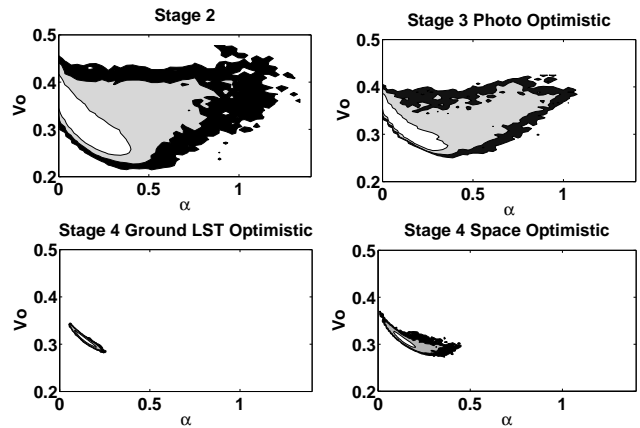}}%
\caption{$V_{0}-\alpha$ $1 \sigma$ $(68.27\%)$, $2 \sigma$ $(95.44\%)$ and $3 \sigma$ $(99.73\%)$ likelihood contours
  for DETF optimistic combined data sets generated from a selected IPL background cosmological model for the case of a cut-off of $log_{10}(\phi_I) = -3$ 
placed on the MCMC algorithm. This effectively gives an enlarged and more detailed view of non-tracking and ``thawing''-like regions
of the parameter space. \label{V0_alpha_opt_iplfid2_b1a}}
\end{figure}

As
indicated previously, in order to allow the MCMC to step more
frequently in non-tracking portions of the parameter space and so
bring out greater  detail in the thawing and some of the transient
portions, we have also generated MCMC chains with a lower bound of
$-3$ placed on $log_{10}(\phi_I)$. These outlying  regions of
parameter space associated with the thawing equation-of-state behavior
(again corresponding to $\phi_I \rightarrow  M_P$ and increasing $w$
and  present-day dark energy density values in recent times) can 
clearly be seen in the Stage 2 and Stage 3 error contours
for our IPL fiducial model  depicted in Fig. \ref{V0_alpha_opt_iplfid2_b1a} 
at $\alpha \agt 0.5$, where the
contours turn or ``flare'' upward and become more ``patchy''. This
un-smooth and  flared appearance of the $2$ and $3 \sigma$ contours
correspond to the largest values of $\phi_I$, where the equation of
state $w(a)$ increases or does not turn down  as steeply near
scale factors of unity and so is exhibiting thawing-like behavior, and,
therefore, the acceptable range of $\alpha$ values significantly
increases  as $w$ increases. 

These portions of the likelihood contours
correspond to outlier points lying relatively far outside the main
distribution of parameter points  stepped to by the MCMC chain. In
these regions of parameter space the scalar field starts to evolve on
the flatter portions of the IPL potential  where $V(\phi_I)$ is
small. We see that there is a greater spread in $V_{0}$ values, and,
thus, $V_{0}$ is less constrained by the data here. This is related to
the  fact, again, that the corresponding equation of state values for
$\phi_I \sim M_P$ don't turn down as steeply near scale factors
of unity (or  even increase towards values greater than $-1$) compared
to $w$ values corresponding to $\phi_I < M_P$. We can see that the
area of 
the these outlying likelihood  contours shrinks and tightens from
Stage 2 to Stage 3 and again from Stage 3 to Stage 4. The reduction in
the $V_{0}$ direction again shows improving constraints with
increasing stage number that the data places on these outlying
transient and thawing regions. We also observe an apparent
illustration here of the ability of the  Stage 4 ground-based
simulated data sets to better constrain the thawing behavior than the
Stage 4 space-based data. This appears to be consistent with the
results  obtained in the MCMC analysis for the tracking regions of
parameter space. Moreover, the Stage 4 space-based data rules out a
significant portion of the thawing  region of the parameter space,
while the Stage 4 Ground LST Optimistic data sets appear to rule out
nearly all of the thawing parameter values.

\begin{figure}[h !]
\scalebox{0.26}{\includegraphics{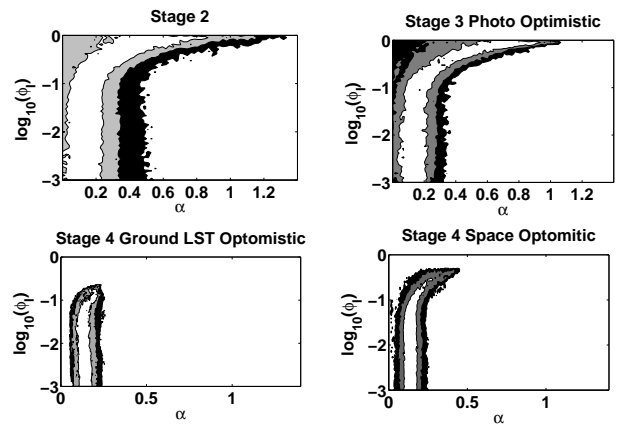}}%
\caption{$log(\phi_I)-\alpha$ $1 \sigma$ $(68.27\%)$, $2 \sigma$ $(95.44\%)$ and $3 \sigma$ $(99.73\%)$ likelihood contours
  for DETF optimistic combined data sets generated from a selected IPL background cosmological model for the case of a cut-off of $log_{10}(\phi_I) = -3$ 
placed on the MCMC algorithm. This effectively gives an enlarged and more detailed view of non-tracking and ``thawing''-like regions
of the parameter space. \label{phiI_alpha_opt_iplfid2_b1a}}
\end{figure}

Figure \ref{phiI_alpha_opt_iplfid2_b1a} depicts $log(\phi_I)-\alpha$
likelihood contours for our IPL fiducial model for non-tracking
regions of parameter space associated with transient and outlying
thawing equation of state behavior. Given that our MCMC analysis did
not focus nearly as much on non-tracking regions of parameter space
than the tracking regions, our chains may not have equilibrated for
the non-tracking regions to the same extent as they have done for
tracking regions. However, we believe that important trends can still
be ascertained from this analysis. In the Stage 2 and Stage 3
likelihood  contours, a significant increase in the $\alpha$ direction
for the largest $\phi_I$ values ($\phi_I \rightarrow M_P$) can be
seen. This corresponds to the fact, as  discussed in Section
\ref{sec:Sec2b}, that for the IPL model the range of acceptable
$\alpha$ values is largest for the largest initial scalar field values
from which the attractor is not joined by the present time. This is
associated with the flatter part of the IPL potential that is less
sensitive to $\alpha$, which controls the slope of the potential. 

As
can be seen from Eq.~(\ref{slope}), flatter parts of the potential
correspond to cases where $V(\phi_I) \sim V(\phi_0)$ is small and
$\phi_I$ is large. Thus, even large values of $\alpha$ can be
associated with flatter portions of the potential here. So, as long as
$V(\phi_I) \sim V(\phi_0)$ remains very small and the ratio of
$\alpha$ and $\phi_I$ does not become too large, a larger range of
acceptable values of $\alpha$, leading to similar cosmologies, will be
allowed within the parameter space. Moreover, larger $\phi_I$ values
combined with larger $\alpha$ can lead to similar $w(a) \agt -1$ with
behavior close to that  of $\Lambda$. Hence, for Stage 3 data and
especially Stage 4 data, the MCMC will not step as much in this region
(since $\Lambda CDM$ models and models with similar  behavior are
ruled out to a greater extent by Stage 3 and 4 data). This explains
the greater constraints placed in the $\alpha$ direction for the very
largest $\phi_I$ values for successive stages of data sets, and is
consistent with the overall trend of the likelihood contours in the
$log(\phi_I)-\alpha$ space shrinking in  the $\alpha$ direction with
higher quality data. 

The extent to which larger $\alpha$ values (and
thus significant portions of the thawing regions of the parameter
space) are constrained and even ruled out by the Stage 4 data sets
also reflects the degree to which evolving dark energy is constrained
and disfavored by the higher  quality data sets. Once again, and
perhaps more dramatically illustrated here, we see that the
ground-based Stage 4 data sets constrain the thawing regions of
parameter space for the IPL model to a more significant extent than do
the Stage 4 space-based data sets. 

\section{\label{sec:Sec5}Discussion and Conclusions}

We have presented our MCMC analysis of the inverse power law
quintessence model using combined simulated data sets forecast by the
DETF and  representing future dark energy experiments. In doing so, we
have analyzed the impact of DETF simulated data models in the context
of the IPL model of dark energy and demonstrated the ability of these
experiments to place significant constraints on the parameters of a
quintessence model. We have  found that the effect of the DETF
combined data models on the parameter space of IPL models is broadly
consistent with the DETF findings. In particular,  we have found a
significant improvement in the constraining power of each successive
stage of DETF simulated data sets. 

We have shown likelihood contours
for choices of combined DETF data sets and found the increase in IPL
dark energy parameter constraints with increasing data quality to be
consistent with the DETF results in the $w_0-w_a$ parameter
space. For example, the relative constraints on the size of the $V_0 -
\phi_I$ parameter space between different simulated data sets lead to
similar constraints computed by the DETF in the $w_0-w_a$ parameter
space. A direct comparison with the DETF Figure of Merit was
complicated by the fact that the IPL model depends on 3 parameters
($\alpha$, $\phi_I$, and $V_{0}$), while the DETF FoM was calculated
based on the two-dimensional $w_0-w_a$ space. However, we found that
the changes in the areas of projected two-dimensional likelihood
contours were consistent with the DETF results. Specifically, the DETF
reported an FoM (defined as the inverse area inside the $95\%$
likelihood contours in the $w_0-w_a$ plane) that showed a gain of at
least a factor of 3 in going from Stage 2 to good combinations of
Stage 3 data sets (and thus a factor of  roughly 3 decrease in allowed
parameter area when moving from Stage 2 to good combinations of Stage
3 data), and a gain of at least a factor of 10 in going from Stage 2
to good combinations of Stage 4 projects. We observed decreases by
similar amounts in our projected 2-D likelihood contours for pairs of
IPL parameters. 

In the course of this work we have also produced and
examined similar 2-D likelihood plots of a much wider range of
combined DETF  simulated data sets, including data models with
``pessimistic'' estimates of systematic errors and data models
representing single DE observing techniques. We found our results in
the IPL model parameter space to be consistent with the constraints
reported by the DETF in the $w_0-w_a$ space across the  complete range
of data combinations and selections that we considered. 

We constructed our simulated data sets from two different background
cosmologies, one with a cosmological constant and one with an IPL
scalar field with specific parameter values. We found our results to
be consistent with those of the DETF in both cases.  We have
separately analyzed cases constrained to having early tracking
behavior and other cases which focused on the non-tracking
solutions. In each case we have placed bounds on some of the IPL potential
parameters as necessary to prevent the MCMC from infinitely stepping
in divergent directions of parameter space (and thus never converging
to a stationary  probability distribution) and to also enable us to
better examine and analyze details in enlarged regions of parameter
space corresponding to non-tracking behavior.  

In order to demonstrate
the power Stage 4 experiments will have for detecting the evolution of
dark energy, we chose  a specific background IPL scalar
field model with parameter values of $\alpha = 0.14$, $\phi_I =
10^{-15}$, and $V_{0} = 0.31$ that was consistent with  Stage 2 data
based on a cosmological constant. This specific model corresponds to 
$w(a=1) \equiv w_0 = -0.95535$, which deviates from $w = -1$ by about
$4.5\%$. One must look back to much earlier times (e.g., $a < 0.2$)
and/or look to larger $\alpha$ parameter values in order to find more
significant  deviation from $w=-1$ for this quintessence model (see
Figs. \ref{fig:fig1ab_ipl}, \ref{fig:ipl_fig2}, \ref{fig:ipl_fig3},
\ref{fig:ipl_fig4}). We found that if the universe were in fact to be
described by this fiducial IPL quintessence model, then good Stage 4 experiments
would rule out a $\Lambda CDM$ model by better than $3 \sigma$,
indicating that there is indeed a
dynamical component to dark energy. For the IPL background cosmology,
we found that the $\Lambda CDM$ model lies outside the  $1 \sigma$
contour but within the $2 \sigma$ contour at Stage 2 and lies outside
of the $2 \sigma$ likelihood contour by Stage 3. We also noted that
the variable  $\alpha$ was somewhat more strongly constrained by Stage
4 ground data sets than with Stage 4 space data. This is consistent
with the results reported by  \cite{barnard2007} for a similar MCMC
analysis carried out on the Albrecht-Skordis scalar field model, but
is opposite of the behavior displayed by the  Exponential and PNGB
scalar field models as described in our other companion papers
\cite{abrahamse2007,bozek2007}. This effect is under current
investigation and may lead to new insights into the complementarity of 
ground and space-based Stage 4 dark energy projects. 
   
We have found, as also discussed in \cite{albrecht2007} and
demonstrated in our companion papers
\cite{abrahamse2007,barnard2007,bozek2007}, that  widely varying
families of functions $w(a)$ for the IPL model are constrained by the
DETF data sets in a similar way to the constraints found in the
$w_{0}-w_{a}$ parameter space by the DETF. In particular, we have seen
that the main IPL model potential parameter $\alpha$ is constrained by
DETF data models in a comparable way to the constraints found in the
$w_{0}-w_{a}$ formulation by the DETF, even though the  $w_{0}-w_{a}$
parameters describe very different functions $w(a)$. We believe that
this relates to the fact pointed out in \cite{albrec2007} that high
quality DETF data sets will be able to constrain many more properties
of $w(a)$ that are present in the $w_{0}-w_{a}$ parameterization alone
and will thus be able to make good  measurements of significantly more
than two equation of state parameters. More specifically, by
considering the IPL family of $w(a)$ 
functions and $w_{0}-w_{a}$ family of functions in terms of an
orthonormal basis of independently measure mode functions $w_{j}(a)$
(as discussed in \cite{albrec2007,albrecht2007}), we are able to
ensure that a wide variety of different $w(a)$ functions will be
constrained as well as the DETF $w_{0}-w_{a}$ parameters. In other
words, the various quintessence models (discussed in this paper and in
our companion papers) are just  sampling different random combinations
of the ``well measured modes'' discussed in \cite{albrec2007} and in
each case lead to similar results. This also appears to reflect the
fact that many more functions $w(a)$ are measured than are contained
in any of the quintessence model $w(a)$ family of  functions alone
\cite{barnard2007}. Consequently, modeling the impact of future dark
energy experiments using the two-parameter DETF scheme makes some
sense in that it gives a good indicator of the impact of scalar field dark energy
models with a similar number of parameters in the quintessence
potential. 

One of the advantages of the techniques
employed in this and the companion work
\cite{abrahamse2007,barnard2007,bozek2007} is that we can explicitly 
examine how simulated data sets representing future dark energy
experiments can constrain actual theoretically motivated
quintessence models (in addition to abstract parameterizations such as
the $w_{0}-w_{a}$ ansatz) in a significant way.  As developed further in
\cite{barnard2008} this approach helps us
understand how future data has the capability to reject some (or
possibly even all) current dark energy models entirely.

\begin{acknowledgments}

We would like to thank David Ring for useful discussions, technical assistance, and 
for finding an error in our code. We also acknowledge Tony Tyson and his group for 
the use of their computer cluster, and, in particular, Perry Gee and Hu Zhan for their expert 
advice and computing support. We also thank Gary Bernstein for providing us with 
Fisher matrices suitable for adapting the DETF weak lensing data models to our methods. 
This work was  supported in part by DOE grant DE-FG03-91ER40674 and NSF grant AST-0632901. 
\end{acknowledgments}
\bibliography{ipl31}

\begin{thebibliography}{44}
\expandafter\ifx\csname natexlab\endcsname\relax\def\natexlab#1{#1}\fi
\expandafter\ifx\csname bibnamefont\endcsname\relax
  \def\bibnamefont#1{#1}\fi
\expandafter\ifx\csname bibfnamefont\endcsname\relax
  \def\bibfnamefont#1{#1}\fi
\expandafter\ifx\csname citenamefont\endcsname\relax
  \def\citenamefont#1{#1}\fi
\expandafter\ifx\csname url\endcsname\relax
  \def\url#1{\texttt{#1}}\fi
\expandafter\ifx\csname urlprefix\endcsname\relax\def\urlprefix{URL }\fi
\providecommand{\bibinfo}[2]{#2}
\providecommand{\eprint}[2][]{\url{#2}}

\bibitem[{\citenamefont{Ratra and Peebles}(1988)}]{ratra88}
\bibinfo{author}{\bibfnamefont{B.}~\bibnamefont{Ratra}} \bibnamefont{and}
  \bibinfo{author}{\bibfnamefont{P.~J.~E.} \bibnamefont{Peebles}},
  \bibinfo{journal}{Phys. Rev. D} \textbf{\bibinfo{volume}{37}},
  \bibinfo{pages}{3406} (\bibinfo{year}{1988}).

\bibitem[{\citenamefont{Albrecht}(2007)}]{albrecht2007}
\bibinfo{author}{\bibfnamefont{A.}~\bibnamefont{Albrecht}},
  \bibinfo{journal}{AIP Conf. Proc.} \textbf{\bibinfo{volume}{957}},
  \bibinfo{pages}{3} (\bibinfo{year}{2007}), \eprint{arXiv:0710.0867}.

\bibitem[{\citenamefont{Bousso}(2008)}]{bousso2008}
\bibinfo{author}{\bibfnamefont{R.}~\bibnamefont{Bousso}},
  \bibinfo{journal}{Gen. Relativ. Gravit.} \textbf{\bibinfo{volume}{40}},
  \bibinfo{pages}{607} (\bibinfo{year}{2008}).

\bibitem[{\citenamefont{Abrahamse et~al.}(2008)\citenamefont{Abrahamse,
  Albrecht, Barnard, and Bozek}}]{abrahamse2007}
\bibinfo{author}{\bibfnamefont{A.}~\bibnamefont{Abrahamse}},
  \bibinfo{author}{\bibfnamefont{A.}~\bibnamefont{Albrecht}},
  \bibinfo{author}{\bibfnamefont{M.}~\bibnamefont{Barnard}}, \bibnamefont{and}
  \bibinfo{author}{\bibfnamefont{B.}~\bibnamefont{Bozek}},
  \bibinfo{journal}{Phys. Rev.} \textbf{\bibinfo{volume}{D77}},
  \bibinfo{pages}{103503} (\bibinfo{year}{2008}).

\bibitem[{\citenamefont{Bozek et~al.}(2008)\citenamefont{Bozek, Abrahamse,
  Albrecht, and Barnard}}]{bozek2007}
\bibinfo{author}{\bibfnamefont{B.}~\bibnamefont{Bozek}},
  \bibinfo{author}{\bibfnamefont{A.}~\bibnamefont{Abrahamse}},
  \bibinfo{author}{\bibfnamefont{A.}~\bibnamefont{Albrecht}}, \bibnamefont{and}
  \bibinfo{author}{\bibfnamefont{M.}~\bibnamefont{Barnard}},
  \bibinfo{journal}{Phys. Rev.} \textbf{\bibinfo{volume}{D77}},
  \bibinfo{pages}{103504} (\bibinfo{year}{2008}).

\bibitem[{\citenamefont{Barnard
  et~al.}(2008{\natexlab{a}})\citenamefont{Barnard, Abrahamse, Albrecht, Bozek,
  and Yashar}}]{barnard2007}
\bibinfo{author}{\bibfnamefont{M.}~\bibnamefont{Barnard}},
  \bibinfo{author}{\bibfnamefont{A.}~\bibnamefont{Abrahamse}},
  \bibinfo{author}{\bibfnamefont{A.}~\bibnamefont{Albrecht}},
  \bibinfo{author}{\bibfnamefont{B.}~\bibnamefont{Bozek}}, \bibnamefont{and}
  \bibinfo{author}{\bibfnamefont{M.}~\bibnamefont{Yashar}},
  \bibinfo{journal}{Phys. Rev.} \textbf{\bibinfo{volume}{D77}},
  \bibinfo{pages}{103502} (\bibinfo{year}{2008}{\natexlab{a}}).

\bibitem[{\citenamefont{Barnard
  et~al.}(2008{\natexlab{b}})\citenamefont{Barnard, Abrahamse, Albrecht, Bozek,
  and Yashar}}]{barnard2008}
\bibinfo{author}{\bibfnamefont{M.}~\bibnamefont{Barnard}},
  \bibinfo{author}{\bibfnamefont{A.}~\bibnamefont{Abrahamse}},
  \bibinfo{author}{\bibfnamefont{A.}~\bibnamefont{Albrecht}},
  \bibinfo{author}{\bibfnamefont{B.}~\bibnamefont{Bozek}}, \bibnamefont{and}
  \bibinfo{author}{\bibfnamefont{M.}~\bibnamefont{Yashar}},
  \bibinfo{journal}{Phys. Rev.} \textbf{\bibinfo{volume}{D78}},
  \bibinfo{pages}{043528} (\bibinfo{year}{2008}{\natexlab{b}}).

\bibitem[{\citenamefont{Albrecht et~al.}(2006)}]{DETF06}
\bibinfo{author}{\bibfnamefont{A.}~\bibnamefont{Albrecht}} \bibnamefont{et~al.}
  (\bibinfo{year}{2006}), \eprint{astro-ph/0609591}.

\bibitem[{\citenamefont{Linder}(2003)}]{linder2002}
\bibinfo{author}{\bibfnamefont{E.~V.} \bibnamefont{Linder}},
  \bibinfo{journal}{Phys. Rev. Lett.} \textbf{\bibinfo{volume}{90}},
  \bibinfo{pages}{091301} (\bibinfo{year}{2003}).

\bibitem[{\citenamefont{Liddle et~al.}(2006)\citenamefont{Liddle, Mukherjee,
  Parkinson, and Wang}}]{liddle2006}
\bibinfo{author}{\bibfnamefont{A.~R.} \bibnamefont{Liddle}},
  \bibinfo{author}{\bibfnamefont{P.}~\bibnamefont{Mukherjee}},
  \bibinfo{author}{\bibfnamefont{D.}~\bibnamefont{Parkinson}},
  \bibnamefont{and} \bibinfo{author}{\bibfnamefont{Y.}~\bibnamefont{Wang}},
  \bibinfo{journal}{Physical Rev. D} \textbf{\bibinfo{volume}{74}},
  \bibinfo{pages}{123506} (\bibinfo{year}{2006}).

\bibitem[{\citenamefont{Copeland et~al.}(2006)\citenamefont{Copeland, Sami, and
  Tsujikawa}}]{copeland2006}
\bibinfo{author}{\bibfnamefont{E.~J.} \bibnamefont{Copeland}},
  \bibinfo{author}{\bibfnamefont{M.}~\bibnamefont{Sami}}, \bibnamefont{and}
  \bibinfo{author}{\bibfnamefont{S.}~\bibnamefont{Tsujikawa}},
  \bibinfo{journal}{Int. J. of Mod. Phys. D} \textbf{\bibinfo{volume}{15}},
  \bibinfo{pages}{1753} (\bibinfo{year}{2006}).

\bibitem[{\citenamefont{Zlatev et~al.}(1999)\citenamefont{Zlatev, Wang, and
  Steinhardt}}]{zlatev1998}
\bibinfo{author}{\bibfnamefont{I.}~\bibnamefont{Zlatev}},
  \bibinfo{author}{\bibfnamefont{L.-M.} \bibnamefont{Wang}}, \bibnamefont{and}
  \bibinfo{author}{\bibfnamefont{P.~J.} \bibnamefont{Steinhardt}},
  \bibinfo{journal}{Phys. Rev. Lett.} \textbf{\bibinfo{volume}{82}},
  \bibinfo{pages}{896} (\bibinfo{year}{1999}).

\bibitem[{\citenamefont{Steinhardt et~al.}(1999)\citenamefont{Steinhardt, Wang,
  and Zlatev}}]{steinhardt-1999-59}
\bibinfo{author}{\bibfnamefont{P.~J.} \bibnamefont{Steinhardt}},
  \bibinfo{author}{\bibfnamefont{L.}~\bibnamefont{Wang}}, \bibnamefont{and}
  \bibinfo{author}{\bibfnamefont{I.}~\bibnamefont{Zlatev}},
  \bibinfo{journal}{Phys. Rev. D} \textbf{\bibinfo{volume}{59}},
  \bibinfo{pages}{123504} (\bibinfo{year}{1999}).

\bibitem[{\citenamefont{Liddle and Scherrer}(1999)}]{liddle1998}
\bibinfo{author}{\bibfnamefont{A.~R.} \bibnamefont{Liddle}} \bibnamefont{and}
  \bibinfo{author}{\bibfnamefont{R.~J.} \bibnamefont{Scherrer}},
  \bibinfo{journal}{Phys. Rev. D} \textbf{\bibinfo{volume}{59}},
  \bibinfo{pages}{023509} (\bibinfo{year}{1999}).

\bibitem[{\citenamefont{Steinhardt}(2005)}]{steinhardt05}
\bibinfo{author}{\bibfnamefont{P.~J.} \bibnamefont{Steinhardt}},
  \bibinfo{journal}{Physica Scripta Volume T} \textbf{\bibinfo{volume}{117}},
  \bibinfo{pages}{34} (\bibinfo{year}{2005}).

\bibitem[{\citenamefont{Masiero et~al.}(1999)\citenamefont{Masiero, Pietroni,
  and Rosati}}]{masiero2000}
\bibinfo{author}{\bibfnamefont{A.}~\bibnamefont{Masiero}},
  \bibinfo{author}{\bibfnamefont{M.}~\bibnamefont{Pietroni}}, \bibnamefont{and}
  \bibinfo{author}{\bibfnamefont{F.}~\bibnamefont{Rosati}},
  \bibinfo{journal}{Phys. Rev. D} \textbf{\bibinfo{volume}{61}},
  \bibinfo{pages}{023504} (\bibinfo{year}{1999}).

\bibitem[{\citenamefont{Peebles and Ratra}(1988)}]{peebles87}
\bibinfo{author}{\bibfnamefont{P.~J.~E.} \bibnamefont{Peebles}}
  \bibnamefont{and} \bibinfo{author}{\bibfnamefont{B.}~\bibnamefont{Ratra}},
  \bibinfo{journal}{Astrophys. J.} \textbf{\bibinfo{volume}{325}},
  \bibinfo{pages}{L17} (\bibinfo{year}{1988}).

\bibitem[{\citenamefont{Martin}(2008)}]{martin2008}
\bibinfo{author}{\bibfnamefont{J.}~\bibnamefont{Martin}}
  (\bibinfo{year}{2008}), \eprint{arXiv:0803.4076}.

\bibitem[{\citenamefont{Caldwell and Doran}(2004)}]{caldwell2004}
\bibinfo{author}{\bibfnamefont{R.~R.} \bibnamefont{Caldwell}} \bibnamefont{and}
  \bibinfo{author}{\bibfnamefont{M.}~\bibnamefont{Doran}},
  \bibinfo{journal}{Phys. Rev. D} \textbf{\bibinfo{volume}{69}},
  \bibinfo{pages}{103517} (\bibinfo{year}{2004}).

\bibitem[{\citenamefont{Eriksson and Amanullah}(2002)}]{eriksson2002}
\bibinfo{author}{\bibfnamefont{M.}~\bibnamefont{Eriksson}} \bibnamefont{and}
  \bibinfo{author}{\bibfnamefont{R.}~\bibnamefont{Amanullah}},
  \bibinfo{journal}{Phys. Rev. D} \textbf{\bibinfo{volume}{66}},
  \bibinfo{pages}{023530} (\bibinfo{year}{2002}).

\bibitem[{\citenamefont{Giovi et~al.}(2003)\citenamefont{Giovi, Baccigalupi,
  and Perrotta}}]{giovi2003}
\bibinfo{author}{\bibfnamefont{F.}~\bibnamefont{Giovi}},
  \bibinfo{author}{\bibfnamefont{C.}~\bibnamefont{Baccigalupi}},
  \bibnamefont{and} \bibinfo{author}{\bibfnamefont{F.}~\bibnamefont{Perrotta}},
  \bibinfo{journal}{Phys. Rev. D} \textbf{\bibinfo{volume}{68}},
  \bibinfo{pages}{123002} (\bibinfo{year}{2003}).

\bibitem[{\citenamefont{Caldwell and Linder}(2005)}]{caldwell2005}
\bibinfo{author}{\bibfnamefont{R.~R.} \bibnamefont{Caldwell}} \bibnamefont{and}
  \bibinfo{author}{\bibfnamefont{E.~V.} \bibnamefont{Linder}},
  \bibinfo{journal}{Phys. Rev. Lett.} \textbf{\bibinfo{volume}{95}},
  \bibinfo{pages}{141301} (\bibinfo{year}{2005}).

\bibitem[{\citenamefont{Malquarti and Liddle}(2002)}]{malquarti2002}
\bibinfo{author}{\bibfnamefont{M.}~\bibnamefont{Malquarti}} \bibnamefont{and}
  \bibinfo{author}{\bibfnamefont{A.~R.} \bibnamefont{Liddle}},
  \bibinfo{journal}{Phys. Rev. D} \textbf{\bibinfo{volume}{66}},
  \bibinfo{pages}{023524} (\bibinfo{year}{2002}).

\bibitem[{\citenamefont{Scherrer}(2006)}]{scherrer2006}
\bibinfo{author}{\bibfnamefont{R.~J.} \bibnamefont{Scherrer}},
  \bibinfo{journal}{Physical Review D} \textbf{\bibinfo{volume}{73}},
  \bibinfo{pages}{043502} (\bibinfo{year}{2006}).

\bibitem[{\citenamefont{Kneller and Strigari}(2003)}]{kneller2003}
\bibinfo{author}{\bibfnamefont{J.~P.} \bibnamefont{Kneller}} \bibnamefont{and}
  \bibinfo{author}{\bibfnamefont{L.~E.} \bibnamefont{Strigari}},
  \bibinfo{journal}{Phys. Rev. D} \textbf{\bibinfo{volume}{68}},
  \bibinfo{pages}{083517} (\bibinfo{year}{2003}).

\bibitem[{\citenamefont{Yahiro et~al.}(2002)\citenamefont{Yahiro, Mathews,
  Ichiki, Kajino, and Orito}}]{yahiro2001}
\bibinfo{author}{\bibfnamefont{M.}~\bibnamefont{Yahiro}},
  \bibinfo{author}{\bibfnamefont{G.~J.} \bibnamefont{Mathews}},
  \bibinfo{author}{\bibfnamefont{K.}~\bibnamefont{Ichiki}},
  \bibinfo{author}{\bibfnamefont{T.}~\bibnamefont{Kajino}}, \bibnamefont{and}
  \bibinfo{author}{\bibfnamefont{M.}~\bibnamefont{Orito}},
  \bibinfo{journal}{Phys. Rev. D} \textbf{\bibinfo{volume}{65}},
  \bibinfo{pages}{063502} (\bibinfo{year}{2002}).

\bibitem[{\citenamefont{Brax and Martin}(2000)}]{brax1999}
\bibinfo{author}{\bibfnamefont{P.}~\bibnamefont{Brax}} \bibnamefont{and}
  \bibinfo{author}{\bibfnamefont{J.}~\bibnamefont{Martin}},
  \bibinfo{journal}{Phys. Rev. D} \textbf{\bibinfo{volume}{61}},
  \bibinfo{pages}{103502} (\bibinfo{year}{2000}).

\bibitem[{\citenamefont{Watson and Scherrer}(2003)}]{watson2003}
\bibinfo{author}{\bibfnamefont{C.~R.} \bibnamefont{Watson}} \bibnamefont{and}
  \bibinfo{author}{\bibfnamefont{R.~J.} \bibnamefont{Scherrer}},
  \bibinfo{journal}{Phys. Rev. D} \textbf{\bibinfo{volume}{68}},
  \bibinfo{pages}{123524} (\bibinfo{year}{2003}).

\bibitem[{\citenamefont{Peebles and Ratra}(2003)}]{peebles2002}
\bibinfo{author}{\bibfnamefont{P.~J.~E.} \bibnamefont{Peebles}}
  \bibnamefont{and} \bibinfo{author}{\bibfnamefont{B.}~\bibnamefont{Ratra}},
  \bibinfo{journal}{Rev. Mod. Phys.} \textbf{\bibinfo{volume}{75}},
  \bibinfo{pages}{559} (\bibinfo{year}{2003}).

\bibitem[{\citenamefont{Astier et~al.}(2006)}]{astier2006}
\bibinfo{author}{\bibfnamefont{P.}~\bibnamefont{Astier}} \bibnamefont{et~al.}
  (\bibinfo{collaboration}{The SNLS}), \bibinfo{journal}{Astron. Astrophys.}
  \textbf{\bibinfo{volume}{447}}, \bibinfo{pages}{31} (\bibinfo{year}{2006}).

\bibitem[{\citenamefont{{Spergel} et~al.}(2007)\citenamefont{{Spergel}, {Bean},
  {Dor{\'e}}, {Nolta}, {Bennett}, {Dunkley}, {Hinshaw}, {Jarosik}, {Komatsu},
  {Page} et~al.}}]{spergel2007}
\bibinfo{author}{\bibfnamefont{D.~N.} \bibnamefont{{Spergel}}},
  \bibinfo{author}{\bibfnamefont{R.}~\bibnamefont{{Bean}}},
  \bibinfo{author}{\bibfnamefont{O.}~\bibnamefont{{Dor{\'e}}}},
  \bibinfo{author}{\bibfnamefont{M.~R.} \bibnamefont{{Nolta}}},
  \bibinfo{author}{\bibfnamefont{C.~L.} \bibnamefont{{Bennett}}},
  \bibinfo{author}{\bibfnamefont{J.}~\bibnamefont{{Dunkley}}},
  \bibinfo{author}{\bibfnamefont{G.}~\bibnamefont{{Hinshaw}}},
  \bibinfo{author}{\bibfnamefont{N.}~\bibnamefont{{Jarosik}}},
  \bibinfo{author}{\bibfnamefont{E.}~\bibnamefont{{Komatsu}}},
  \bibinfo{author}{\bibfnamefont{L.}~\bibnamefont{{Page}}},
  \bibnamefont{et~al.}, \bibinfo{journal}{Astrophys. J. Suppl.}
  \textbf{\bibinfo{volume}{170}}, \bibinfo{pages}{377} (\bibinfo{year}{2007}).

\bibitem[{\citenamefont{Riess et~al.}(2007)}]{riess2007}
\bibinfo{author}{\bibfnamefont{A.~G.} \bibnamefont{Riess}}
  \bibnamefont{et~al.}, \bibinfo{journal}{Astrophys. J.}
  \textbf{\bibinfo{volume}{659}}, \bibinfo{pages}{98} (\bibinfo{year}{2007}).

\bibitem[{\citenamefont{Kowalski et~al.}(2008)}]{kowalski2008}
\bibinfo{author}{\bibfnamefont{M.}~\bibnamefont{Kowalski}}
  \bibnamefont{et~al.}, \bibinfo{journal}{Astrophys. J.}
  \textbf{\bibinfo{volume}{686}}, \bibinfo{pages}{749} (\bibinfo{year}{2008}).

\bibitem[{\citenamefont{Baccigalupi et~al.}(2002)\citenamefont{Baccigalupi,
  Balbi, Matarrese, Perrotta, and Vittorio}}]{baccigalupi2001}
\bibinfo{author}{\bibfnamefont{C.}~\bibnamefont{Baccigalupi}},
  \bibinfo{author}{\bibfnamefont{A.}~\bibnamefont{Balbi}},
  \bibinfo{author}{\bibfnamefont{S.}~\bibnamefont{Matarrese}},
  \bibinfo{author}{\bibfnamefont{F.}~\bibnamefont{Perrotta}}, \bibnamefont{and}
  \bibinfo{author}{\bibfnamefont{N.}~\bibnamefont{Vittorio}},
  \bibinfo{journal}{Phys. Rev. D} \textbf{\bibinfo{volume}{65}},
  \bibinfo{pages}{063520} (\bibinfo{year}{2002}).

\bibitem[{\citenamefont{Amendola and Quercellini}(2003)}]{amendola2003}
\bibinfo{author}{\bibfnamefont{L.}~\bibnamefont{Amendola}} \bibnamefont{and}
  \bibinfo{author}{\bibfnamefont{C.}~\bibnamefont{Quercellini}},
  \bibinfo{journal}{Phys. Rev. D} \textbf{\bibinfo{volume}{68}},
  \bibinfo{pages}{023514} (\bibinfo{year}{2003}).

\bibitem[{\citenamefont{Colombo and Gervasi}(2006)}]{colombo2006}
\bibinfo{author}{\bibfnamefont{L.~P.~L.} \bibnamefont{Colombo}}
  \bibnamefont{and} \bibinfo{author}{\bibfnamefont{M.}~\bibnamefont{Gervasi}},
  \bibinfo{journal}{JCAP} \textbf{\bibinfo{volume}{0610}}, \bibinfo{pages}{001}
  (\bibinfo{year}{2006}).

\bibitem[{\citenamefont{Barro~Calvo and Maroto}(2006)}]{calvo2006}
\bibinfo{author}{\bibfnamefont{G.}~\bibnamefont{Barro~Calvo}} \bibnamefont{and}
  \bibinfo{author}{\bibfnamefont{A.~L.} \bibnamefont{Maroto}},
  \bibinfo{journal}{Phys. Rev. D} \textbf{\bibinfo{volume}{74}},
  \bibinfo{pages}{083519} (\bibinfo{year}{2006}).

\bibitem[{\citenamefont{Schimd et~al.}(2007)}]{schimd2006}
\bibinfo{author}{\bibfnamefont{C.}~\bibnamefont{Schimd}} \bibnamefont{et~al.},
  \bibinfo{journal}{Astron. Astrophys.} \textbf{\bibinfo{volume}{463}},
  \bibinfo{pages}{405} (\bibinfo{year}{2007}).

\bibitem[{\citenamefont{Schimd and Tereno}(2007)}]{schimd2006a}
\bibinfo{author}{\bibfnamefont{C.}~\bibnamefont{Schimd}} \bibnamefont{and}
  \bibinfo{author}{\bibfnamefont{I.}~\bibnamefont{Tereno}},
  \bibinfo{journal}{J. Phys. A} \textbf{\bibinfo{volume}{40}},
  \bibinfo{pages}{7105} (\bibinfo{year}{2007}).

\bibitem[{\citenamefont{Efstathiou}(2008)}]{efstathiou2008}
\bibinfo{author}{\bibfnamefont{G.}~\bibnamefont{Efstathiou}},
  \bibinfo{journal}{Monthly Notices of the Royal Astronomical Society}
  \textbf{\bibinfo{volume}{388}} (\bibinfo{year}{2008}).

\bibitem[{\citenamefont{Bludman}(2004)}]{bludman2004}
\bibinfo{author}{\bibfnamefont{S.}~\bibnamefont{Bludman}},
  \bibinfo{journal}{Phys. Rev. D} \textbf{\bibinfo{volume}{69}},
  \bibinfo{pages}{122002} (\bibinfo{year}{2004}).

\bibitem[{\citenamefont{de~la Macorra and
  Stephan-Otto}(2002)}]{delaMacorra2001}
\bibinfo{author}{\bibfnamefont{A.}~\bibnamefont{de~la Macorra}}
  \bibnamefont{and}
  \bibinfo{author}{\bibfnamefont{C.}~\bibnamefont{Stephan-Otto}},
  \bibinfo{journal}{Phys. Rev. D} \textbf{\bibinfo{volume}{65}},
  \bibinfo{pages}{083520} (\bibinfo{year}{2002}).

\bibitem[{\citenamefont{Albrecht and Bernstein}(2007)}]{albrec2007}
\bibinfo{author}{\bibfnamefont{A.}~\bibnamefont{Albrecht}} \bibnamefont{and}
  \bibinfo{author}{\bibfnamefont{G.}~\bibnamefont{Bernstein}},
  \bibinfo{journal}{Phys. Rev. D} \textbf{\bibinfo{volume}{75}},
  \bibinfo{pages}{103003} (\bibinfo{year}{2007}).

\bibitem[{\citenamefont{Freedman et~al.}(2001)}]{Freedman2000}
\bibinfo{author}{\bibfnamefont{W.~L.} \bibnamefont{Freedman}}
  \bibnamefont{et~al.}, \bibinfo{journal}{Astrophys. J.}
  \textbf{\bibinfo{volume}{553}}, \bibinfo{pages}{47} (\bibinfo{year}{2001}).

\end{thebibliography}
\end{document}